\documentclass[fleqn,usenatbib]{mnras}

\usepackage{newtxtext,newtxmath}

\usepackage[T1]{fontenc}

\DeclareRobustCommand{\VAN}[3]{#2}
\let\VANthebibliography\thebibliography
\def\thebibliography{\DeclareRobustCommand{\VAN}[3]{##3}\VANthebibliography}

\usepackage[english]{babel}
\usepackage{hyperref}
\addto\extrasenglish{
    
}
\addto\extrasenglish{
    
}

\usepackage{graphicx}	
\usepackage{amsmath}
\title[Thick Disk Planetary System TOI-2345]{An Ultra-Short Period Super-Earth and Sub-Neptune Spanning the Radius Valley Orbiting the Kinematic Thick Disk Star TOI-2345}

\author[Y. N. E. Eschen et al.]{Yoshi~Nike~Emilia~Eschen$^{1}$\,$^{\href{https://orcid.org/0009-0006-6397-2503}{\protect\includegraphics[height=0.19cm]{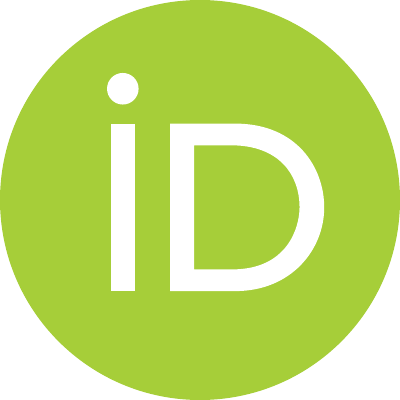}}}$, 
Thomas~G.~Wilson$^{1}$\,$^{\href{https://orcid.org/0000-0001-8749-1962}{\protect\includegraphics[height=0.19cm]{figures/orcid.pdf}}}$, 
Andrea~Bonfanti$^{2}$\,$^{\href{https://orcid.org/0000-0002-1916-5935}{\protect\includegraphics[height=0.19cm]{figures/orcid.pdf}}}$, 
Carina~M.~Persson$^{3}$, \newauthor
Sérgio~G.~Sousa$^{4}$\,$^{\href{https://orcid.org/0000-0001-9047-2965}{\protect\includegraphics[height=0.19cm]{figures/orcid.pdf}}}$, 
Monika~Lendl$^{5}$\,$^{\href{https://orcid.org/0000-0001-9699-1459}{\protect\includegraphics[height=0.19cm]{figures/orcid.pdf}}}$, 
Alexis~Heitzmann$^{5}$\,$^{\href{https://orcid.org/0000-0002-8091-7526}{\protect\includegraphics[height=0.19cm]{figures/orcid.pdf}}}$, 
Attila~E.~Simon$^{6,7}$\,$^{\href{https://orcid.org/0000-0001-9773-2600}{\protect\includegraphics[height=0.19cm]{figures/orcid.pdf}}}$, 
Göran~Olofsson$^{8}$\,$^{\href{https://orcid.org/0000-0003-3747-7120}{\protect\includegraphics[height=0.19cm]{figures/orcid.pdf}}}$, \newauthor
Amadeo~Castro-González$^{5}$, 
Jo~Ann~Egger$^{6}$\,$^{\href{https://orcid.org/0000-0003-1628-4231}{\protect\includegraphics[height=0.19cm]{figures/orcid.pdf}}}$, 
Luca~Fossati$^{2}$\,$^{\href{https://orcid.org/0000-0003-4426-9530}{\protect\includegraphics[height=0.19cm]{figures/orcid.pdf}}}$, 
Alexander~James~Mustill$^{9}$\,$^{\href{https://orcid.org/0000-0002-2086-3642}{\protect\includegraphics[height=0.19cm]{figures/orcid.pdf}}}$, \newauthor
Hugh~P.~Osborn$^{7,10}$\,$^{\href{https://orcid.org/0000-0002-4047-4724}{\protect\includegraphics[height=0.19cm]{figures/orcid.pdf}}}$, 
Hugo~G.~Vivien$^{11}$\,$^{\href{https://orcid.org/0000-0001-7239-6700}{\protect\includegraphics[height=0.19cm]{figures/orcid.pdf}}}$, 
Yann~Alibert$^{7,6}$\,$^{\href{https://orcid.org/0000-0002-4644-8818}{\protect\includegraphics[height=0.19cm]{figures/orcid.pdf}}}$, 
Roi~Alonso$^{12,13}$\,$^{\href{https://orcid.org/0000-0001-8462-8126}{\protect\includegraphics[height=0.19cm]{figures/orcid.pdf}}}$, 
Tamas~Bárczy$^{14}$\,$^{\href{https://orcid.org/0000-0002-7822-4413}{\protect\includegraphics[height=0.19cm]{figures/orcid.pdf}}}$, \newauthor
David~Barrado$^{15}$\,$^{\href{https://orcid.org/0000-0002-5971-9242}{\protect\includegraphics[height=0.19cm]{figures/orcid.pdf}}}$, 
Susana~C.~C.~Barros$^{4,16}$\,$^{\href{https://orcid.org/0000-0003-2434-3625}{\protect\includegraphics[height=0.19cm]{figures/orcid.pdf}}}$, 
Wolfgang~Baumjohann$^{2}$\,$^{\href{https://orcid.org/0000-0001-6271-0110}{\protect\includegraphics[height=0.19cm]{figures/orcid.pdf}}}$, 
Willy~Benz$^{6,7}$\,$^{\href{https://orcid.org/0000-0001-7896-6479}{\protect\includegraphics[height=0.19cm]{figures/orcid.pdf}}}$, \newauthor
Nicolas~Billot$^{5}$\,$^{\href{https://orcid.org/0000-0003-3429-3836}{\protect\includegraphics[height=0.19cm]{figures/orcid.pdf}}}$, 
Luca~Borsato$^{17}$\,$^{\href{https://orcid.org/0000-0003-0066-9268}{\protect\includegraphics[height=0.19cm]{figures/orcid.pdf}}}$, 
Alexis~Brandeker$^{8}$\,$^{\href{https://orcid.org/0000-0002-7201-7536}{\protect\includegraphics[height=0.19cm]{figures/orcid.pdf}}}$, 
Christopher~Broeg$^{6,7}$\,$^{\href{https://orcid.org/0000-0001-5132-2614}{\protect\includegraphics[height=0.19cm]{figures/orcid.pdf}}}$, 
Maximilian~Buder$^{18}$, \newauthor
Douglas~A.~~Caldwell$^{19}$\,$^{\href{https://orcid.org/0000-0003-1963-9616}{\protect\includegraphics[height=0.19cm]{figures/orcid.pdf}}}$, 
Andrew~Collier~Cameron$^{20}$\,$^{\href{https://orcid.org/0000-0002-8863-7828}{\protect\includegraphics[height=0.19cm]{figures/orcid.pdf}}}$, 
Alexandre~C.~M.~Correia$^{21}$\,$^{\href{https://orcid.org/0000-0002-8946-8579}{\protect\includegraphics[height=0.19cm]{figures/orcid.pdf}}}$, \newauthor
Szilard~Csizmadia$^{18}$\,$^{\href{https://orcid.org/0000-0001-6803-9698}{\protect\includegraphics[height=0.19cm]{figures/orcid.pdf}}}$, 
Patricio~E.~Cubillos$^{2,22}$, 
Melvyn~B.~Davies$^{23}$\,$^{\href{https://orcid.org/0000-0001-6080-1190}{\protect\includegraphics[height=0.19cm]{figures/orcid.pdf}}}$, 
Magali~Deleuil$^{11}$\,$^{\href{https://orcid.org/0000-0001-6036-0225}{\protect\includegraphics[height=0.19cm]{figures/orcid.pdf}}}$, 
Adrien~Deline$^{5}$, \newauthor
Olivier~D.~S.~Demangeon$^{4,16}$\,$^{\href{https://orcid.org/0000-0001-7918-0355}{\protect\includegraphics[height=0.19cm]{figures/orcid.pdf}}}$, 
Brice-Olivier~Demory$^{7,24,6}$\,$^{\href{https://orcid.org/0000-0002-9355-5165}{\protect\includegraphics[height=0.19cm]{figures/orcid.pdf}}}$, 
Aliz~Derekas$^{25}$, 
Billy~Edwards$^{26}$, \newauthor
David~Ehrenreich$^{5,27}$\,$^{\href{https://orcid.org/0000-0001-9704-5405}{\protect\includegraphics[height=0.19cm]{figures/orcid.pdf}}}$, 
Anders~Erikson$^{18}$, 
Jacopo~Farinato$^{17}$\,$^{\href{https://orcid.org/0000-0002-5840-8362}{\protect\includegraphics[height=0.19cm]{figures/orcid.pdf}}}$, 
Andrea~Fortier$^{6,7}$\,$^{\href{https://orcid.org/0000-0001-8450-3374}{\protect\includegraphics[height=0.19cm]{figures/orcid.pdf}}}$, \newauthor
Malcolm~Fridlund$^{28,3}$\,$^{\href{https://orcid.org/0000-0002-0855-8426}{\protect\includegraphics[height=0.19cm]{figures/orcid.pdf}}}$, 
Davide~Gandolfi$^{29}$\,$^{\href{https://orcid.org/0000-0001-8627-9628}{\protect\includegraphics[height=0.19cm]{figures/orcid.pdf}}}$, 
Kosmas~Gazeas$^{30}$\,$^{\href{https://orcid.org/0000-0002-8855-3923}{\protect\includegraphics[height=0.19cm]{figures/orcid.pdf}}}$, 
Michaël~Gillon$^{31}$\,$^{\href{https://orcid.org/0000-0003-1462-7739}{\protect\includegraphics[height=0.19cm]{figures/orcid.pdf}}}$, 
Robert~Goeke$^{32}$, \newauthor
Manuel~Güdel$^{33}$, 
Maximilian~N.~Günther$^{34}$\,$^{\href{https://orcid.org/0000-0002-3164-9086}{\protect\includegraphics[height=0.19cm]{figures/orcid.pdf}}}$, 
Johann~Hasiba$^{2}$, 
Ch.~Helling$^{2,35}$, 
Kate~G.~Isaak$^{34}$\,$^{\href{https://orcid.org/0000-0001-8585-1717}{\protect\includegraphics[height=0.19cm]{figures/orcid.pdf}}}$, \newauthor
Jon~M.~~Jenkins$^{36}$, 
Tatiana~Keller$^{6,7}$, 
Laszlo~L.~Kiss$^{37,38}$, 
Daniel~Kitzmann$^{6,7}$, 
Judith~Korth$^{5}$\,$^{\href{https://orcid.org/0000-0002-0076-6239}{\protect\includegraphics[height=0.19cm]{figures/orcid.pdf}}}$, \newauthor
Kristine~W.~F.~Lam$^{18}$\,$^{\href{https://orcid.org/0000-0002-9910-6088}{\protect\includegraphics[height=0.19cm]{figures/orcid.pdf}}}$, 
Jacques~Laskar$^{39}$\,$^{\href{https://orcid.org/0000-0003-2634-789X}{\protect\includegraphics[height=0.19cm]{figures/orcid.pdf}}}$, 
Alain~Lecavelier~des~Etangs$^{40}$\,$^{\href{https://orcid.org/0000-0002-5637-5253}{\protect\includegraphics[height=0.19cm]{figures/orcid.pdf}}}$, 
Adrien~Leleu$^{5,6}$\,$^{\href{https://orcid.org/0000-0003-2051-7974}{\protect\includegraphics[height=0.19cm]{figures/orcid.pdf}}}$, \newauthor
Demetrio~Magrin$^{17}$\,$^{\href{https://orcid.org/0000-0003-0312-313X}{\protect\includegraphics[height=0.19cm]{figures/orcid.pdf}}}$, 
Pierre~F.~L.~Maxted$^{41}$\,$^{\href{https://orcid.org/0000-0003-3794-1317}{\protect\includegraphics[height=0.19cm]{figures/orcid.pdf}}}$, 
Bruno~Merín$^{42}$\,$^{\href{https://orcid.org/0000-0002-8555-3012}{\protect\includegraphics[height=0.19cm]{figures/orcid.pdf}}}$, 
Christoph~Mordasini$^{6,7}$, \newauthor
Valerio~Nascimbeni$^{17}$\,$^{\href{https://orcid.org/0000-0001-9770-1214}{\protect\includegraphics[height=0.19cm]{figures/orcid.pdf}}}$, 
Roland~Ottensamer$^{33}$, 
Isabella~Pagano$^{43}$\,$^{\href{https://orcid.org/0000-0001-9573-4928}{\protect\includegraphics[height=0.19cm]{figures/orcid.pdf}}}$, 
Enric~Pallé$^{12,13}$\,$^{\href{https://orcid.org/0000-0003-0987-1593}{\protect\includegraphics[height=0.19cm]{figures/orcid.pdf}}}$, 
Gisbert~Peter$^{18}$\,$^{\href{https://orcid.org/0000-0001-6101-2513}{\protect\includegraphics[height=0.19cm]{figures/orcid.pdf}}}$, \newauthor
Daniele~Piazza$^{44}$, 
Giampaolo~Piotto$^{17,45}$\,$^{\href{https://orcid.org/0000-0002-9937-6387}{\protect\includegraphics[height=0.19cm]{figures/orcid.pdf}}}$, 
Don~Pollacco$^{1}$, 
Didier~Queloz$^{10,46}$\,$^{\href{https://orcid.org/0000-0002-3012-0316}{\protect\includegraphics[height=0.19cm]{figures/orcid.pdf}}}$, 
Roberto~Ragazzoni$^{17,45}$\,$^{\href{https://orcid.org/0000-0002-7697-5555}{\protect\includegraphics[height=0.19cm]{figures/orcid.pdf}}}$, \newauthor
Nicola~Rando$^{34}$, 
Francesco~Ratti$^{34}$, 
Heike~Rauer$^{47,48}$\,$^{\href{https://orcid.org/0000-0002-6510-1828}{\protect\includegraphics[height=0.19cm]{figures/orcid.pdf}}}$, 
Ignasi~Ribas$^{49,50}$\,$^{\href{https://orcid.org/0000-0002-6689-0312}{\protect\includegraphics[height=0.19cm]{figures/orcid.pdf}}}$, 
Nuno~C.~Santos$^{4,16}$\,$^{\href{https://orcid.org/0000-0003-4422-2919}{\protect\includegraphics[height=0.19cm]{figures/orcid.pdf}}}$, \newauthor 
Gaetano~Scandariato$^{43}$\,$^{\href{https://orcid.org/0000-0003-2029-0626}{\protect\includegraphics[height=0.19cm]{figures/orcid.pdf}}}$, 
Damien~Ségransan$^{5}$\,$^{\href{https://orcid.org/0000-0003-2355-8034}{\protect\includegraphics[height=0.19cm]{figures/orcid.pdf}}}$, 
Avi~Shporer$^{32}$, 
Alexis~M.~S.~Smith$^{18}$\,$^{\href{https://orcid.org/0000-0002-2386-4341}{\protect\includegraphics[height=0.19cm]{figures/orcid.pdf}}}$, \newauthor
Manu~Stalport$^{51,31}$, 
Sophia~Sulis$^{11}$\,$^{\href{https://orcid.org/0000-0001-8783-526X}{\protect\includegraphics[height=0.19cm]{figures/orcid.pdf}}}$, 
Gyula~M.~Szabó$^{25,52}$\,$^{\href{https://orcid.org/0000-0002-0606-7930}{\protect\includegraphics[height=0.19cm]{figures/orcid.pdf}}}$, 
Stéphane~Udry$^{5}$\,$^{\href{https://orcid.org/0000-0001-7576-6236}{\protect\includegraphics[height=0.19cm]{figures/orcid.pdf}}}$, 
Solène~Ulmer-Moll$^{53,51}$\,$^{\href{https://orcid.org/0000-0003-2417-7006}{\protect\includegraphics[height=0.19cm]{figures/orcid.pdf}}}$,\newauthor 
Valérie~Van~Grootel$^{51}$\,$^{\href{https://orcid.org/0000-0003-2144-4316}{\protect\includegraphics[height=0.19cm]{figures/orcid.pdf}}}$, 
Julia~Venturini$^{5}$\,$^{\href{https://orcid.org/0000-0001-9527-2903}{\protect\includegraphics[height=0.19cm]{figures/orcid.pdf}}}$, 
Eva~Villaver$^{12,13}$, 
Nicholas~A.~Walton$^{54}$\,$^{\href{https://orcid.org/0000-0003-3983-8778}{\protect\includegraphics[height=0.19cm]{figures/orcid.pdf}}}$, 
David~Watanabe$^{55}$, \newauthor
Sebastian~Wolf$^{6}$, and
Carl~Ziegler$^{56}$
\\
Affiliations are listed at the end of the paper
}

\date{Accepted XXX. Received YYY; in original form ZZZ}

\pubyear{\the\year{}}

\newcommand{\steffsme}[1][]{$4687 \pm 60$} 
\newcommand{\sloggsme}{$4.57\pm 0.06$} 
\newcommand{\sfehsme}[1][]{$-0.10\pm 0.07$} 
\newcommand{\smghsme}[1][]{$0.02 \pm 0.11$} 
\newcommand{\ssihsme}[1][]{$-0.12 \pm 0.09$} 
\newcommand{\svsinisme}[1][]{$2.3 \pm 0.9$} 
\newcommand{\svmic}[1][]{$0.1$} 
\newcommand{\svmac}[1][]{$1.0$} 

\begin{document}
\label{firstpage}
\pagerange{\pageref{firstpage}--\pageref{lastpage}}
\maketitle

\begin{abstract}
A crucial chemical link between stars and their orbiting exoplanets is thought to exist. If universal, this connection could affect the formation and evolution of all planets. Therefore, this potential vital link needs testing by characterising exoplanets around chemically-diverse stars. We present the discovery of two planets orbiting the metal-poor, kinematic thick-disk K-dwarf TOI-2345. TOI-2345\,b is a super-Earth with a period of 1.05 days and TOI-2345\,c is a sub-Neptune with a period of 21 days. In addition to the target being observed in 4 \textit{TESS} sectors, we obtained 5 \textit{CHEOPS} visits and 26 radial velocities from HARPS. By conducting a joint analysis of all the data, we find TOI-2345\,b to have a radius of $1.504\substack{+0.047\\-0.044}$\,R$_\oplus$ and a mass of $3.49\pm0.85$\,M$_\oplus$; and TOI-2345\,c to have a radius of $2.451\substack{+0.045\\-0.046}$\,R$_\oplus$ and a mass of $7.27\substack{+2.27\\-2.45}$\,M$_\oplus$. To explore chemical links between these planets and their host star, we model their interior structures newly accounting for devolatised stellar abundances. TOI-2345 adds to the limited sample of well characterised planetary systems around thick disk stars. This system challenges theories of formation and populations of planets around thick disk stars with its Ultra-Short Period super-Earth and the wide period distribution of these two planets spanning the radius valley. 
\end{abstract}




\section{Introduction}
Since the discovery of the first exoplanet orbiting a main sequence star, 51 Peg \citep{Mayor_Queloz_51Peg_1995}, the field has been exponentially growing. Beginning with radial velocity surveys in the late 1990s and early 2000s \citep[e.g. ][]{Queloz_GJ86_2000,Santos_mu_Arae_2004} a few hundreds of exoplanets were detected, the majority of which are of high masses such as Jupiter. During the last decades instruments such as HARPS \citep{Pepe_HARPS_2000,Mayor_HARPS_2003}, ESPRESSO \citep{Pepe_ESPRESSO_2021} or EXPRES \citep{Jurgenson_EXPRES_2016} obtaining higher precisions, as well as our ability to model stellar activity, has opened the door to measure smaller mass planets down to that of Earth \citep[e.g. ][]{Gonzales_Hernandez_Barnards_Star_2024,Turner_Gliese12_2025}. However, the main increase in the number of exoplanet detections over the last 15 years was due to space-based photometric missions detecting transiting planets and deriving their radii. NASA's \textit{Kepler} mission \citep{Borucki_Kepler_2010} observed a fixed field in the Northern Hemisphere from 2009 to 2013. This mission discovered nearly 3000 confirmed and validated planets \citep[][accessed on 17 July 2025]{Exoplanet_Archive}. After the failure of two of the four reaction wheels the \textit{Kepler} mission was continued as \textit{K2} \citep{Howell_K2_2014} observing several fields around the ecliptic until 2018, adding to the sample of transiting planets. Since 2018, NASA's Transiting Exoplanet Survey Satellite \citep[\textit{TESS};][]{Ricker_TESS_2015} has been performing an all-sky survey and has detected 643 planets to date \citep[][accessed on 17 July 2025]{Exoplanet_Archive}. Additionally, ESA's photometric mission, CHaracterising ExOPlanet Satellite \citep[\textit{CHEOPS};][]{Benz_CHEOPS_2021}, is following up discovered transiting planets and planet candidates to derive their radii more precisely. By the end of 2026, ESA's upcoming PLAnetary Transits and Oscillations of stars mission \citep[\textit{PLATO};][]{Rauer_PLATO_2025} is expected to launch. As this mission will observe bright and nearby targets for at least two years and with up to 24 cameras, it is expected to reach higher precisions than \textit{TESS} and \textit{CHEOPS} allowing to better characterise transiting planets and discover further systems. \textit{PLATO}'s long continuous observation will also enables finding planets of longer orbital periods including outer planets in known systems with short period planets \citep{Eschen_PLATO_Sensitivities_2025,Rauer_PLATO_2025}.

Among these newly discovered planets by \textit{Kepler} were two planet types not found in our own Solar System: super-Earths and sub-Neptunes \citep{Batalha_Kepler_Candidates_2013}. The large sample of \textit{Kepler} planets allowed demographics studies with these planet types found to be the most common in our Milky Way \citep{Howard_Kepler_Planet_Occurrence_2012,Petigura_super_earths_plateau_2013}. By studying their radius distribution, a dearth of planets between $\sim$1.5-2\,R$_\oplus$ was found \citep{Fulton_Radius_Valley_2017,Remo_Radius_Valley_2024}, named the radius valley. For FGK dwarfs the radius valley seems to arise from atmospheric mass loss \citep{VanEylen_Radius_Valley_2018}. In addition to atmospheric escape from either photo-evaporation or core-powered mass loss \citep{Owen_Schlichting_Photoevaporation_2024} the radius valley can also be explained by formation and evolution models \citep{Venturini_Radius_Valley_2020}. Importantly, the radius valley is stellar mass and orbital period dependent \citep{Ho_Radius_Valley_Kepler_2023}. However, not many multi-planetary systems that span the valley, especially at long orbital periods, have been studied. Hence these systems remain exciting to study and provide insights into planet formation and evolution processes.  

Below the radius valley lie super-Earths. Within this sample, a population of Ultra-Short Period (USP) planets have been discovered \citep{Sanchis_Ojeda_USP_Kepler_2014,Adams_USP_K2_2021}. These are planets that orbit their host star in less than or roughly 1\,d \citep{Goyal_USP_2025} and are hence highly irradiated. Therefore they have likely lost their atmospheres and can provide insights into the deeper interior of small planets \citep{Winn_USP_2017,Dai_Hot_Earths_2019}. Since several hundreds of these planets have been found to date \citep[][accessed on 17 July 2025]{Exoplanet_Archive}, they have been part of several demographics studies \citep{Winn_Kepler78_USP_2018}. These include the findings of \citet{Dai_USP_inclinations_2018} reporting that USPs have higher mutual inclinations than other systems. \citet{Tu_USP_Age_Architecture_2025} found that USPs are more often found around older thick disk stars and the period spacings between them and outer bodies seem to increase with age. Since their origin and evolution is still not fully understood they remain interesting planets to characterise. Overall, the origin and evolution of USPs is not fully understood, even though it is commonly thought that these planets migrated inwards through interactions with outer companions \citep[e.g.][]{Petrovich_USP_high_eccentricity_2019,Pu_Lai_USP_low_eccentricity_2019,Millholland_USP_Obliquity_2020}. 

Since the planetary radius and orbital inclination can be derived from the photometric data and the mass multiplied by the inclination from radial velocity observations, combining the two methods is valuable to characterise systems well. Knowing the bulk density of small planets (in this context R$_\text{P}$<4\,R$_\oplus$) allows modelling of their interior structure \citep{Huang_MAGRATHEA_2022,Baumeister_ExoMDN_2023,Egger_TOI469_plaNETic_2024}. This provides insight into planet formation and evolution mechanisms such as core accretion and atmospheric escape \citep{Owen_Murray_Clay_Metallicity_Connection_Kepler_2018,Armitage2020,Kubyshkina_Fossati_thermal_evolution_hydrodynamic_escape_2022}. As found by \citet{Adibekyan_compositional_link_2021,Wilson_TOI1064_2022,Adibekyan_compositional_link_2024} there may be a compositional link between the host star abundances and the interior structure of these planets since they are formed from the same material and the abundances are unlikely to have changed significantly during the formation \citep{Thiabaud_elements_stars_planets_2015,Nielsen_Planet_Formation_2023,Steffen_Galactic_Chemical_Evolution_2025} as seen by the abundances of refractory elements in the proto-Sun and Earth \citep{Wang_Protosun_Abundances_2019}. 

However, there is a lack of planets around metal-poor and $\alpha$-enhanced stars which is not clear if it arises from physical or observational origin. Hence to identify trends linking stellar and planetary composition, and perform demographic studies, this sample needs to be increased. Metal-poor and $\alpha$-enhanced stars are most likely to be found in the kinematic thick disk of the Milky Way \citep{Fuhrmann_Disk_Halo_1998,Hayden_Metallicity_Milky_Way_2015}. This is due to the Interstellar Medium (ISM) being enriched in $\alpha$-elements due to more massive stars (M$>$8M$_\odot$) exploding in type II supernova at the earlier stages of the Milky Way. With more time low-mass stars evolved into white dwarfs and type Ia supernova could enrich the ISM with iron-peak elements \citep{Wheeler_Abundance_1989,Gondoin_chemical_abundances_history_milkyway_2024,Steffen_Galactic_Chemical_Evolution_2025}. Hence stars in the thin disk are found to be metal-rich while stars in the thick disk mainly remain metal-poor \citep{Hayden_Metallicity_Milky_Way_2015}. Hence, to add to the sample of small transiting planets around metal-poor stars, we characterise the planets orbiting the kinematic thick disk star TOI-2345 hosting an USP and a sub-Neptune spanning the radius valley.

We present the data we took with \textit{TESS}, \textit{CHEOPS} and HARPS to discover and characterise this system as well as the ground-based photometry and imaging in \autoref{sec:Observations}, characterise the host star properties in \autoref{sec:Stellar_Characterisation}, model the planetary parameters in \autoref{sec:Planet_Fitting} and discuss its internal structure, atmospheric evolution and place it in the context of other thick disk stars orbited by super-Earths and sub-Neptunes in \autoref{sec:Discussion}. Finally, we conclude in \autoref{sec:Conclusion}.

\section{Observations}
\label{sec:Observations}
In order to characterise the TOI-2345 system, it \textcolor{black}{was} observed with photometric surveys including \textit{TESS}, \textit{CHEOPS}, ASAS-SN and WASP; spectroscopy from HARPS and imaging observations from SOAR. 

\subsection{\textit{TESS}}
The transiting exoplanet survey satellite \citep[\textit{TESS};][]{Ricker_TESS_2015} has been performing an all-sky survey since 2018. \textit{TESS} consists of four cameras each containing four CCDs. To cover the entire sky it divided it up into sectors which are each observed for 27.4\,days. Since its launch, it has identified 7655 planet candidates and discovered 643 new planets \citep[][accessed on 17 July 2025]{Exoplanet_Archive}. Being at the end of its second extended mission now, \textit{TESS} has changed the cadence at which it is observing targets from the original 30 mins in the prime mission, 10 mins during the first extension, and finally the current 200s cadence. \textcolor{black}{Additionally, TESS observes selected targets at a cadence of 2\,min during the primary mission and 20\,s during the extensions.}

\textit{TESS} data is downlinked every 13.7\,days. The data are processed by the Science Processing Operations Centre \citep[SPOC;][]{Jenkins_TESS_SPOC_2016,Caldwell_TESS_SPOC_2020} into lightcurves following the procedures of the \textit{Kepler} pipeline \citep{Jenkins_Kepler_Pipeline_2010}. The produced lightcurves contain flux values obtained from Simple Aperture Photometry \citep[SAP;][]{Twicken_Kepler_Pipeline_2010, Morris_Kepler_Data_Processing_Handbook_2020} as well as the Pre-search Data Conditioning SAP \citep[PDCSAP;][]{Twicken_PDCSAP_2010,Smith_PDCSAP_Kepler_2012, Stumpe_PDCSAP_2012}, which is the SAP flux detrended using Co-trending Basis Vectors and hence showing less systematic trends. In these lightcurves the combined differential photometric precision (CDPP) over 2\,hours is reported \citep{Christiansen_CDPP_2012}. 

Produced lightcurves by SPOC go through a several staged vetting process described in \citet{Guerrero_TOI_Catalogue_2021}. First, a search for transit like signals is conducted. Signals that occur twice or more and have a statistical significance of 7.1$\sigma$ or more as well as some statistical tests pass this stage and are called Threshold Crossing Events (TCEs). To these a transit model is fitted and a summary report including several diagnostic tests gets produced. Within this step the pipeline also searched for further transits in the lightcurve. These data are then passed on to an automated Triage, \textit{TESS}-ExoClass (TEC)\footnote{https://github.com/christopherburke/TESS-ExoClass} based on \textit{Kepler}'s Robovetter \citep{Coughlin_Robovetter_2016,Thompson_Kepler_Robovetter_2018}. Finally targets, passing these tests get vetted manually by a team of Vetters going through the produced reports. Additionally, several teams have developed independent searching and vetting tools \citep[e.g.][]{Montalto_DIAmante_2020,Montalto_DIAmante_2023,Feliz_Nemesis_2021,Olmschenk_Neural_Networks_PCs_2021,Eschen_Vetting_PC_MDwarfs_2024,Kunimoto_LEOVetter_2025}

\begin{figure*}
    \centering    \includegraphics[width=\linewidth]{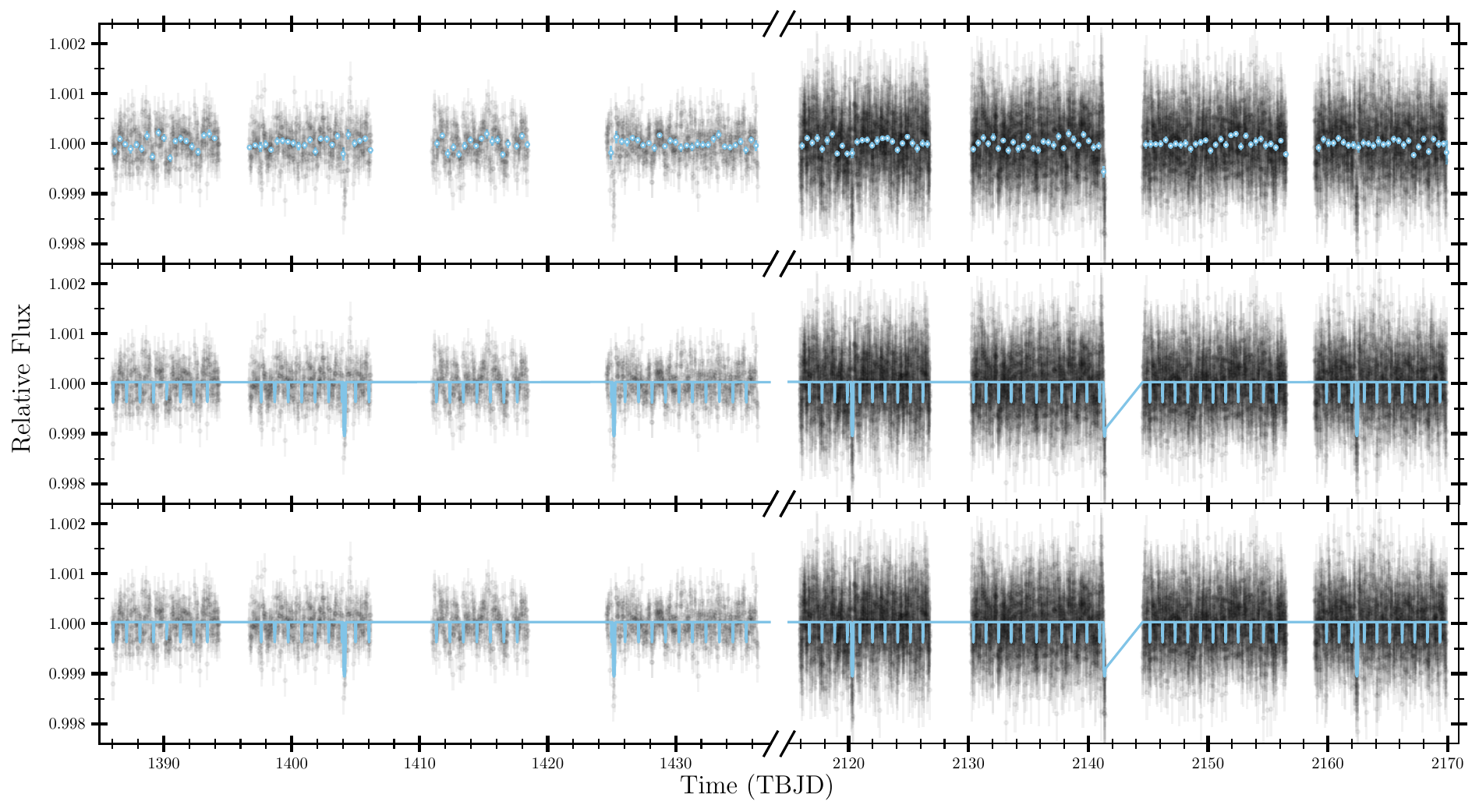}
    \caption{\textit{TESS} data of TOI-2345 in sectors 3, 4, 30 and 31. Top: \textit{TESS} data binned at 10\,hours. Middle: Transit and GP model in blue plotted on top of the data points in gray. Bottom: Transit model and \textit{TESS} data points with the GP model subtracted.}
    \label{fig:tess}
\end{figure*}

In this vetting process, two transiting planet candidates around TOI-2345, TOI-2345.01 and TOI-2345.02, were alerted. Hereafter, we refer to these planets as TOI-2345\,b and c.
TOI-2345 was observed during \textit{TESS}'s Primary mission in sectors 3 and 4 (20 September 2018 to 14 November 2018) with a cadence of 30\,min. The target was observed again in the first Extended Mission within which \textit{TESS} collected data with a cadence of 10\,minutes in sectors 30 and 31 (23 September 2020 to 16 November 2020). \textit{TESS} will re-observe TOI-2345 in October 2025 and Summer 2026 according to \textsc{TESS-point} \citep{Burke_tess_point_2020}. We show and summarise the details of the currently available \textit{TESS} observations in \autoref{fig:tess} and \autoref{tab:tess_obs} respectively.

To perform our own analysis on this target, we downloaded the \textit{TESS} SPOC High-level-science product lightcurves, for sectors 3, 4, 30 and 31. We removed bad quality data (QUALITY>0) points and analysed the PDCSAP flux in this study.\\

\begin{table*}
    \centering
    \caption{\textit{TESS} Observations of TOI-2345.}
    \begin{tabular}{ccccccccc}
        \hline
        \hline
        Sector & Camera & CCD & Start Date & End Date & Cadence & 2-hour CDPP & \#Transits of  & \#Transits \\
         &  &  & (UTC) & (UTC) & (s) & (ppm) & Planet b &  Planet c \\
        \hline
        3 & 2 & 2 & 2018-09-20T13:04:29.861 & 2018-10-17T21:05:07.754 & 1800 & 181.3 & 17 & 1\\
        4 & 2 & 1 & 2018-10-19T10:05:08.066 & 2018-11-14T08:04:33.593 & 1800 & 191.2 & 18 & 1\\
        30 & 2 & 2 & 2020-09-23T09:24:09.487 & 2020-10-20T14:34:38.274 & 600 & 200.2 & 21 & 2\\
        31 & 2 & 1 & 2020-10-22T00:24:38.474 & 2020-11-16T10:43:57.592 & 600 & 184.6 & 21 & 1\\
        \hline
        \hline
    \end{tabular}
    \label{tab:tess_obs}
\end{table*}

\subsection{\textit{CHEOPS}}
The CHaracterising ExOPlanet Satellite \citep[\textit{CHEOPS};][]{Benz_CHEOPS_2021} is a S-class ESA mission launched in 2019. One of its goals is to follow-up known planets in order to derive their radii more precisely which supports constraining planet formation and evolution theories. While \textit{TESS} produces Full Frame Images and makes these publicly available, \textit{CHEOPS} produces window images which contain the observed target. Due to the nadir-locked, Sun-synchronous, low-Earth orbit of \textit{CHEOPS}, each visit contains gaps from Earth occultation that is represented by a observational efficiency. We obtained a total of five \textit{CHEOPS} visits within the \textit{CHEOPS} X-Gal programme (ID: PR120054, PI: Wilson) within the Guaranteed Time Observing programme, see \autoref{fig:cheops_detrending_plots}. These visits cover four transits of TOI-2345\,b and one transit of TOI-2345\,c and are summarised in \autoref{tab:cheops_obs}.

In addition to the aperture photometry from the \textit{CHEOPS} Data Reduction Pipeline \citep[DRP;][]{Hoyer_CHEOPS_DRP_2020}, where we select the optimal aperture per visit based on the lowest RMS value, we used the PSF Imagette Photometric Extraction \citep[PIPE;][]{Brandeker_PIPE_2024} to re-extract the \textit{CHEOPS} photometry using PSF photometry. We fit each visit individually in \textsc{pycheops} \citep{Maxted_pycheops_2023} using \textsc{lmfit} \citep{Newville_lmfit_2014} and the parameters obtained by a \textit{TESS} only fit in \textsc{juliet} \citep[][See \autoref{sec:photometry}]{Espinoza_juliet_2019}. Within \textsc{pycheops} subsets detrending vectors are fitted simultaneously with the transit model. By assessing the Bayes Factor of models containing different combinations of detrending vectors, we are able to obtain the detrending vectors of each visit. We report these selected detrending vectors in \autoref{tab:CHEOPS_detrending}. We use the suggested detrending vectors as linear regressors in \textcolor{black}{ \textsc{juliet}} in order to decorrelate the five \textit{CHEOPS} visits. We apply a 3$\sigma$ clipping, removing outliers that are further away than 3$\sigma$ from the mean of the data. This leaves us with two sets of five detrended \textit{CHEOPS} visits each, one using aperture photometry from the DRP and one using PSF photometry from PIPE.

\begin{table*}
    \centering
    \caption{\textit{CHEOPS} Observations of TOI-2345.}
    \begin{tabular}{cccccccc}
    \hline
    \hline
        Visit & Planets & Start Date & Duration & Data Points & File Key & Efficiency & Exp Time \\
        &  & (UTC) & (h) & (\#) &  & (\%) & (s)\\
        \hline
        1 & b & 2022-10-16T12:30:42 & 12.67 & 529 & CH\_PR120054\_TG001001 & 71.5 & 60\\
        2 & c & 2022-10-25T16:21:42 & 17.29 & 771 & CH\_PR120054\_TG002401 & 74.2 & 60\\
        3 & b & 2022-11-05T13:31:42 & 13.97 & 674 & CH\_PR120054\_TG001002 & 80.3 & 60\\
        4 & b & 2022-11-20T09:50:42 & 11.44 & 428 & CH\_PR120054\_TG003201 & 62.3 & 60\\
        5 & b & 2022-11-21T12:37:43 & 12.41 & 448 & CH\_PR120054\_TG003202 & 60.1 & 60\\
    \hline
    \hline
    \end{tabular}
    \label{tab:cheops_obs}
\end{table*}

\subsection{HARPS}
\label{subsec:HARPS}
We collected 26 high-resolution spectra of TOI-2345 with the High Accuracy Radial Velocity Planet Searcher \citep[HARPS;][]{Pepe_HARPS_2000}. HARPS is a high-resolution Echelle spectrograph mounted on the ESO 3.6\,m telescope in La Silla. HARPS has a wavelength range from 380\,nm to 680\,nm and a resolving power of 90,000. These observations were taken between 2023/07/01 and 2025/01/30 \textcolor{black}{which we show in \autoref{fig:rv_timeseries}}. The typical SNR of these observations at Order 50 is 33.44. These observations were taken as part of the 111.254R program (PI: Wilson) with an exposure time of 1800\,s. The data were processed with the HARPS Data Reduction Software \citep[DRS 3.2.5][]{Lovis_Pepe_HARPS_DRS_2007}. Within the DRS the radial velocities (RVs) as well as activity indicators such as the FWHM, BIS and contrast are computed using the Cross-Correlation function with a K2 mask. Additionally further activity indicators, including the S, H$\alpha$, Na and Ca\,II indices, are computed from the spectra.

As shown by \citet{Silva_sbart_2022}, the precision of RVs can be improved by extracting them through template matching.  Hence we additionally derive RVs using Semi-Bayesian Approach for RVs with Template-matching \citep[\textsc{s-bart};][]{Silva_sbart_2022}. We re-extracted the RVs using the 2D spectra from the DRS and a combination of different template-matching parameters and quality checks and compare the results to find the best combination of these. The template matching fitting parameters include RV steps of 0.1, 0.5, 1.0\,m/s, RV limits of 200, 500, 1000\,m/s and the classical and Laplacian method \textcolor{black}{\textsc{s-bart}} applies. As quality checks we apply minimum order SNRs of 1.5, 5 and 10, airmasses of 1.5, 2.0, 2.2 and 2.6 and RV errors of 5, 6, 7 and 10\,m/s. We obtain a median error of 2.04\,m/s and an RMS of 3.57\,m/s when taking the median of all \textcolor{black}{\textsc{s-bart}} time series median errors and RMS. This is lower than the median error and RMS obtained from the RVs of the DRS which are 3.12\,m/s and 4.61\,m/s respectively. Since \textcolor{black}{\textsc{s-bart}} is reducing the uncertainties, we use it in the further analysis in \autoref{sec:Planet_Fitting}. We describe how we select the optimal RV time series produced by \textcolor{black}{\textsc{s-bart}} from different template matching and quality parameters in \autoref{subsec:Radial_Velocity_Fitting}.




\subsection{ASAS-SN}
The All-Sky Automated Survey for SuperNovae \citep[ASAS-SN;][]{Shappee_ASAS_SN_2014,Kochanek_ASAS_SN_2017} is photometrically monitoring the entire sky to detect transients. As this survey also monitor stars over a long time span it can be used to monitor stellar activity. TOI-2345 was observed in the \textcolor{black}{ASAS-SN V-band from January 2012 to September 2018 and }the g-band by ASAS-SN from September 2017 to August 2025. In the \textcolor{black}{13} years, \textcolor{black}{1062 and} 5497 data points were collected \textcolor{black}{ for the V- and g-Band respectively}. We apply a magnitude cut-off at V=15\,mag and g=15\,mag and 5$\sigma$ clipping to this data, which removes \textcolor{black}{6 and }280 data point, leaving us with \textcolor{black}{1056} and 5217 measurements \textcolor{black}{ respectively}.  

\subsection{WASP}
The Wide Angle Search for Planets \citep[WASP;][]{Pollaco_WASP_2006} has been monitoring stars since 2004 in the Northern and Southern hemisphere to search for transiting planets. This resulted in 170 planet discoveries, of which the majority are Hot Jupiters as these have deep transits. WASP is not precise enough to detect planets as small as Earth, however its long term monitoring can be used to identify the rotation period of stars. TOI-2345 was monitored with WASP from 17 June 2006 to 19 December 2014. During this time WASP collected 12235 data points. We remove data points with relative magnitude errors above 0.01, leaving us with 10324 data points. Applying a 5$\sigma$ clipping to the remaining data points, removes an additional two, leaving 10322 measurements for further analysis.

\subsection{Imaging}
\begin{figure}
    \centering
    \includegraphics[width=\linewidth]{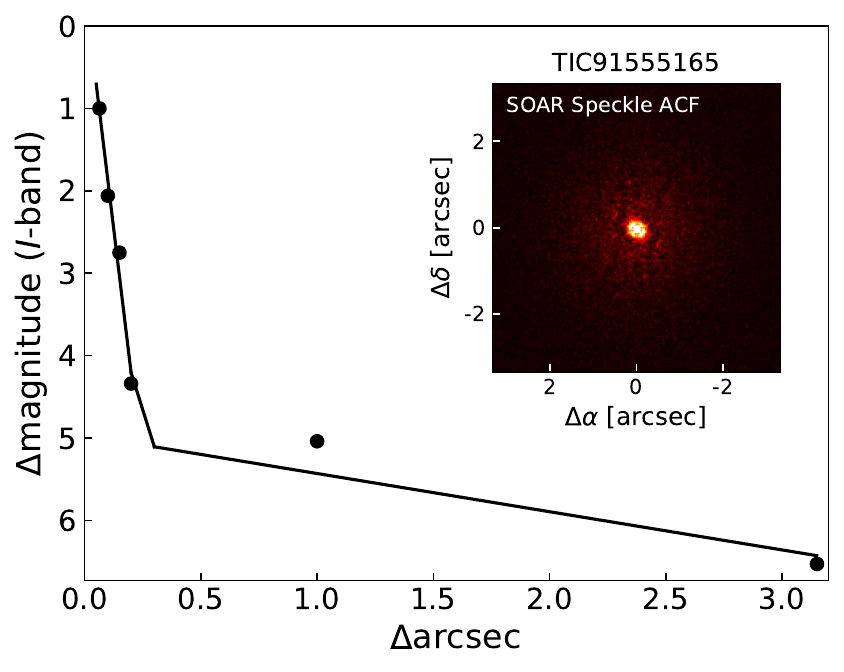}
    \caption{5-$\sigma$ detection sensitivity of the SOAR I-band observation of TOI-2345.}
    \label{fig:imaging}
\end{figure}

High-angular resolution imaging is needed to search for nearby sources that can contaminate the \textit{TESS} photometry, resulting in an underestimated planetary radius, or be the source of astrophysical false positives, such as background eclipsing binaries. We searched for stellar companions to TOI-2345 with speckle imaging on the 4.1-m Southern Astrophysical Research (SOAR) telescope \citep{Tokovinin_SOAR_2018} on 3 December 2020 UT, observing in Cousins I-band, a similar visible bandpass as \textit{TESS}. This observation was sensitive with 5-sigma detection to a 5.0-magnitude fainter star at an angular distance of 1\,arcsec from the target. More details of the observations within the SOAR \textit{TESS} survey are available in \citet{Ziegler_SOAR_TESS_2020}. The 5\,$\sigma$ detection sensitivity and speckle auto-correlation functions from the observations are shown in \autoref{fig:imaging}. No nearby stars were firmly detected within 3\,arcsec of TOI-2345 in the SOAR observations.

\section{Stellar Characterisation}
\label{sec:Stellar_Characterisation}

\begin{table}
\caption{Stellar properties of TOI-2345.}
\begin{center}
\begin{tabular}{lll} 
\hline\hline                 
\multicolumn{3}{c}{TOI-2345} \\    
\hline                        
2MASS & \multicolumn{2}{l}{J02553208-3458391} \\
\textit{Gaia} DR3 & \multicolumn{2}{l}{5049575943053753088} \\
TIC & \multicolumn{2}{l}{91555165} \\
LP & \multicolumn{2}{l}{942-63} \\ 
\hline
Parameter & Value & Note \\ 
\hline
   $\alpha$ [J2000] & 02$^{\rm h}$55$^{\rm m}$32.10$^{\rm s}$ & 1 \\
   $\delta$ [J2000] & -34$^{\circ}$58$^{'}$39.09$^{\arcsec}$ & 1 \\
   $\mu_{\alpha}$ [mas/yr] & 202.481$\pm$0.011 & 1 \\
   $\mu_{\delta}$ [mas/yr] & -104.354$\pm$0.014 & 1 \\
   $\varpi$ [mas] & 12.297$\pm$0.015 & 1 \\
   $d$ [pc] & 81.27$\pm$0.30 & 1 \\
   RV [km~s$^{-1}$] & 58.14$\pm$0.21 & 1 \\
   U [km~s$^{-1}$] & \textcolor{black}{-29.972$\pm$0.056} & 5$^{\rm a}$ \\
   V [km~s$^{-1}$] & \textcolor{black}{-100.23$\pm$0.12} & 5$^{\rm a}$ \\
   W [km~s$^{-1}$] & \textcolor{black}{-13.10$\pm$-0.19} & 5$^{\rm a}$ \\
\hline
   $V$ [mag] & 11.48$\pm$0.08 & 2 \\
   $G_{\rm BP}$ [mag] & 11.599$\pm$0.003 & 1 \\
   $G$ [mag] & 11.030$\pm$0.003 & 1 \\
   $G_{\rm RP}$ [mag] & 10.316$\pm$0.004 & 1 \\
   $J$ [mag] & 9.51$\pm$0.03 & 3 \\
   $H$ [mag] & 8.94$\pm$0.06 & 3 \\
   $K$ [mag] & 8.85$\pm$0.02 & 3 \\
   $W1$ [mag] & 8.76$\pm$0.02 & 4 \\
   $W2$ [mag] & 8.84$\pm$0.02 & 4 \\
\hline
   $T_{\mathrm{eff}}$ [K] & \steffsme & 5; spectroscopy \\
   $\log{g}$ [cm~s$^{-2}$]      & \sloggsme & 5; spectroscopy \\\relax
   [Fe/H] [dex] & \sfehsme[] & 5; spectroscopy \\\relax
   [Mg/H] [dex] & \smghsme & 5; spectroscopy \\\relax
   [Si/H] [dex] & \ssihsme & 5; spectroscopy \\\relax
   $V \sin i_\star$ [km\,s$^{-1}$] & \svsinisme & 5; spectroscopy \\\relax
   $R_{\star}$ [$R_{\odot}$] & 0.729$\pm$0.007 & 5; IRFM \\
   $M_{\star}$ [$M_{\odot}$] & 0.727$\pm$0.033 & 5; isochrones \\  
   $t_{\star}$ [Gyr]        & 6.3$\pm$4.7 & 5; isochrones \\
   $L_{\star}$ [$L_{\odot}$] & 0.231$\pm$0.013 & 5; from $R_{\star}$ and $T_{\mathrm{eff}}$\\
   $\rho_{\star}$ [$\rho_\odot$] & 1.88$\pm$0.10 & 5; from $R_{\star}$ and $M_{\star}$ \\
   $\rho_{\star}$ [$\mathrm{kg\,m^{-3}}$] & 2645$\pm$142 & 5; from $R_{\star}$ and $M_{\star}$ \\
\hline\hline                                   
\end{tabular}
\label{tab:stellarParam}
\end{center}
[1] \cite{Gaia_DR3_Release_Summary_2023}, [2] \cite{Hog2000}, [3] \cite{Skrutskie2006}, [4] \cite{Wright2010}, [5] This work \\
$^{\rm a}$ Calculated via the right-handed, heliocentric Galactic spatial velocity formulation of \cite{Johnson_Soderblom_galactic_space_velocity_1987} using the proper motions, parallax, and radial velocity from [1].
\end{table}

\subsection{Spectral Analysis}
We analysed the co-added high-resolution HARPS spectrum with the software Spectroscopy Made Easy\footnote{\url{http://www.stsci.edu/~valenti/sme.html}} 
 \citep[{\tt SME};][]{vp96, pv2017} to obtain the stellar effective temperature  ($T_\mathrm{eff}$), surface gravity (log\,$g_\star$), and  abundances  ([Fe/H], [Mg/H], [Si/H]). This software  fits observations to computed synthetic spectra based on a chosen stellar atmosphere grid  \citep[Atlas12;][]{Kurucz2013}  and  atomic and molecular line data from
VALD  \citep{Ryabchikova2015}. 
We fitted one parameter at a time following \citet{Persson_K2_216_2018}. We fixed the micro- and macro-turbulent velocities to \svmic~km\,s$^{-1}$ and \svmac~km\,s$^{-1}$ and fitted a large number of iron lines to obtain the projected
equatorial rotational velocity of the star ($V \sin i_\star$).  The parameters derived are $T_{\mathrm{eff}} = 4687 \pm 60$ K, $\log g = 4.57 \pm 0.06$ dex, and [Fe/H] = $ -0.10 \pm 0.07$ dex. 

An additional independent spectral analysis was done with ARES+MOOG as described in \citet{Santos2013,Sousa2014,Sousa2021}. We used the latest version of ARES\footnote{The latest version, ARES v2, can be downloaded at \url{https://github.com/sousasag/ARES}} \citep{Sousa2007, Sousa2015} to consistently measured the equivalent widths (EW) for the iron line list presented in \citet{Sousa2008}. The best spectroscopic parameters are found using the ionization and excitation equilibrium. In this process, it is used for a grid of Kurucz model atmospheres \citep{Kurucz1993} and the radiative transfer code MOOG \citep{Sneden1973}. The parameters derived ($T_{\mathrm{eff}} = 4669 \pm 122$ K, $\log g = 4.55 \pm 0.07$ dex, and [Fe/H] = $ -0.11 \pm 0.05$ dex) are very consistent with the adopted values derived by SME.

\subsection{Radius, Mass, and Age}
\label{subsec:stellar_radius_mass_age}
We determined the stellar radius of TOI-2345 using a MCMC modified infrared flux method \citep[IRFM --][]{Blackwell1977,Schanche2020}. Within this framework we constructed spectral energy distributions (SED) from two stellar atmospheric models catalogues \citep{Kurucz1993,Castelli2003} constrained by our spectroscopically derived stellar parameters. To obtain the stellar bolometric flux, synthetic photometry was produced by the SEDs and compared to observed fluxes in the following bandpasses:  2MASS $J$, $H$, and $K$, WISE $W1$ and $W2$, and \textit{Gaia} $G$, $G_\mathrm{BP}$, and $G_\mathrm{RP}$ \citep{Skrutskie2006,Wright2010,Gaia_DR3_Release_Summary_2023}. From the bolometric flux, we derived the effective temperature and angular diameter that was converted into the stellar radius using the offset-corrected \textit{Gaia} parallax \citep{Lindegren2021}. To account for stellar atmosphere model uncertainties, we took a Bayesian Model Averaging of the stellar radius posterior distributions from the two catalogues. This is reported in \autoref{tab:stellarParam}.

We then inputted the stellar effective temperature $T_{\mathrm{eff}}$, metallicity [Fe/H], and radius $R_{\star}$ along with their uncertainties in the isochrone placement routine \citep{bonfanti2015,bonfanti2016} to derive the stellar mass $M_{\star}$ and age $t_{\star}$ from evolutionary models. Following interpolation within pre-computed grids of PARSEC\footnote{\textsl{PA}dova and T\textsl{R}ieste \textsl{S}tellar \textsl{E}volutionary \textsl{C}ode: \url{https://stev.oapd.inaf.it/cgi-bin/cmd}} v1.2S \citep{marigo2017} isochrones and tracks we obtained $M_{\star}=0.727\pm0.033\,M_{\odot}$ and $t_{\star}=6.3\pm4.7$ Gyr. All the stellar parameters are listed in \autoref{tab:stellarParam}.

\subsection{Rotation Period}
\label{subsec:rotation_period}
\begin{figure*}
    \centering
    \includegraphics[width=\linewidth]{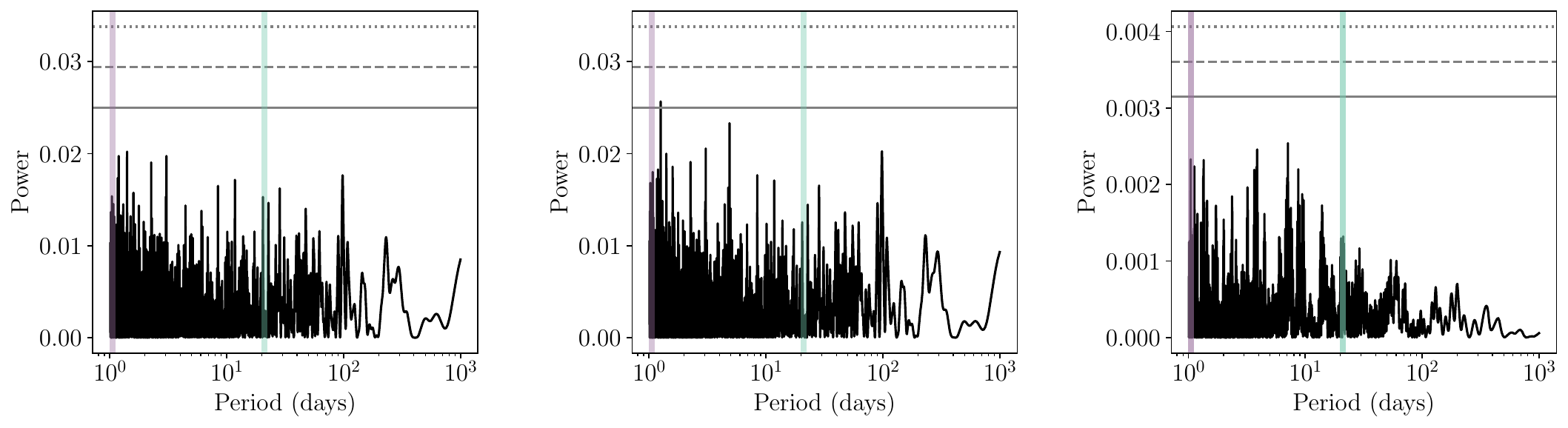}  
    \caption{Lomb-Scargle periodogram of TOI-2345 \textcolor{black}{of the ASAS-SN V-band (left), ASAS-SN g-band (middle) and WASP (right) data}. The orbital periods of the inner and outer planet are highlighted in purple and green respectively. False Alarm Probabilities of 1\%, 0.1\% and 0.01\% are shown by the gray continuous, dashed and dotted line respectively.}
    \label{fig:rotation_period}
\end{figure*}

Since ASAS-SN and WASP cover a long baseline of photometric data as shown in \autoref{fig:asas_sn_wasp_lightcurves}, they can be used to identify stellar rotation periods \citep[e.g.][]{Wilson_TOI1064_2022,Turner_Gliese12_2025}. We hence run a Lomb-Scargle periodogram \citep[][]{Lomb_1976,Scargle_1982} on each dataset individually, \textcolor{black}{dividing the ASAS-SN data into two sets to cover the V- and g-band observations respectively. We remove peaks due to the cadence of the observations\textcolor{black}{, season length} and the moon.} The \textcolor{black}{three} resulting periodograms for ASAS-SN and WASP respectively are shown in \autoref{fig:rotation_period}. The two orbital periods of the planets are highlighted in purple and green. \textcolor{black}{Although TESS covers a shorter baseline than these two surveys, we run a Lomb-Scargle periodogram on the consecutive sectors 3 and 4 a well as 30 and 31 using the SAP flux \textcolor{black}{and remove the transit signals using \textsc{wotan} \citep{Hippke_wotan_2019}.} The underlying data as well as the two periodograms are shown in \autoref{fig:tess_sap} and \autoref{fig:tess_periodogram}.} In \textcolor{black}{all} periodograms, we do not find any significant peaks, concluding that TOI-2345 is inactive from these data. 

\subsection{Kinematic Analysis}
\label{sec:toi2345_kinematics}
\textit{Gaia} DR3 \citep[][]{Gaia_DR3_Release_Summary_2023} reports stellar kinematic properties including the position, parallax, RV and proper motion of stars. The stellar properties of TOI-2345 measured by \textit{Gaia} are reported in \autoref{tab:stellarParam}.
\citet{Johnson_Soderblom_galactic_space_velocity_1987} derive transformations of coordinates and velocities using the equatorial and galactic coordinates of the star as well as its parallax, radial velocity and proper motion. Based on the coordinates of the North Galactic Pole and the position angle of the North Celestial Pole, they compute a transformation matrix. Multiplying this matrix with a coordinate matrix constructed for the star and its radial velocity and proper motion the components of the galactic space-velocity can be computed. Performing this computation for TOI-2345, we obtain (U,V,W)=($-$29.972$\pm$0.056, \textcolor{black}{$-$100.23$\pm$0.12, $-$13.10$\pm$0.19})\,km/s in this heliocentric frame. Since this computation is using a right-handed coordinate system, U is positive in the direction of the Galactic Centre, V is positive towards the Galactic rotation and W is positive towards the North Galactic Pole. Since the galactic space-velocities derived previously are heliocentric, we correct them for the solar motion in the local standards of rest from \citet{Koval_lsr_2009,Coskunoglu_RAVE_LSR_2011,Bensby_lsr_2014,Francis_LSR_2014,Schoenrich_lsr_2010,Almeida_Fernandes_Stellar_Ages_LSR_2018,Tian_Stellar_Kinematic_LAMOST_LSR_2015}.

To compute the galactic kinematic probabilities of the thin disk, thick disk, and halo we follow \citet{Bensby_lsr_2003}. They assume that the stellar populations follow Gaussian distributions which are normalised by the characteristic velocity dispersions of each group ($\sigma_U$,$\sigma_V$,$\sigma_W$) and V is corrected using the asymmetric drift. Since the local number densities of each population are different they multiply the probabilities by the observed fraction of each population to obtain the relative likelihood of a star belonging to either population. \citet{Bensby_lsr_2003} report values for the velocity dispersions, stellar fractions and asymmetric drift of each of these populations. However, since their work several other studies have reported their own values which slightly vary. The studies we used in our analysis are \citet{Bensby_lsr_2003,Reddy_Thick_Disk_2006,Bensby_lsr_2014,Chen_PAST_I_2021}. 
Computing the thick disk probability with all combinations of the local standards of rest and the different velocity dispersions, stellar fractions and asymmetric drifts, we obtained a weighted thick disk probability of 85\%. From this kinematic analysis, we conclude that TOI-2345 is in the Milky Way's thick disk. Due to the cool nature of the host star, we are unable to confirm its place in the thick disk chemically.

\section{Planet Fitting}
\label{sec:Planet_Fitting}

\begin{figure}
    \centering
    \includegraphics[width=\linewidth]{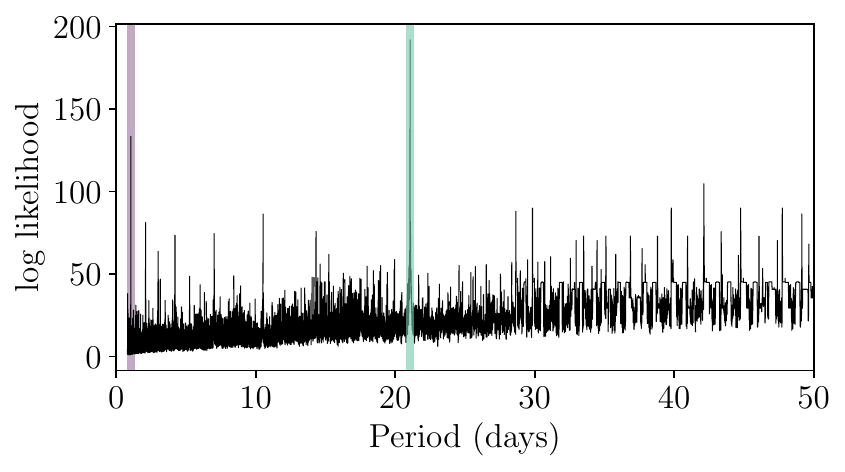}
    \caption{BLS periodogram of the \textit{TESS} lightcurves of TOI-2345. The two significant peaks are highlighted in purple and green.}
    \label{fig:bls}
\end{figure}
To understand the planetary properties, we used \textcolor{black}{\textsc{juliet}} \citep{Espinoza_juliet_2019}, a joint fitting tool using \textcolor{black}{\textsc{batman}} \citep{Kreidberg_batman_2015}  for photometric data and \textcolor{black}{\textsc{radvel}} \citep{Fulton_radvel_2018} to fit the radial velocity data. We make use of \textcolor{black}{\textsc{juliet}} to fit the photometric and radial velocity data individually and jointly. \textcolor{black}{\textsc{Juliet}}  uses \textcolor{black}{\textsc{dynesty}}  \citep{Speagle_dynesty_2020} for nested sampling and computes Bayesian evidences allowing model comparisons. We derive the planetary radii and masses by converting the fitted planet-to-star radius ratios and \textcolor{black}{RV} semi-amplitudes fitted in the following and the radius and mass of the star as reported in \autoref{subsec:stellar_radius_mass_age}
\begin{table*}
        \centering
        \caption{Fitted and Derived Planetary Parameters. Uniform distributions are noted by $\mathcal{U}$.}
        \begin{tabular}{lclrlr}
        \hline
        \hline
             &  & \multicolumn{2}{c}{Planet b} &  \multicolumn{2}{c}{Planet c}\\
             \hline
             Parameter & Unit & prior & posterior & prior & posterior\\
             \hline
             T0 & (BJD) & $\mathcal{U}(2459116.63,2459116.83)$ &  $2459116.7208\substack{+0.0011\\-0.0020}$ & $\mathcal{U}(2459120.20,2459120.40)$ &  $2459120.3007\substack{+0.0014\\-0.0013}$ \\
             P & (days) & $\mathcal{U}(0.95,1.15)$ & $1.0528573\substack{+0.0000025\\-0.0000026}$ & $\mathcal{U}(20.96,21.16)$ & $21.064302\substack{+0.000041\\-0.000041}$ \\
             R$_\text{P}$/R$_*$ & & $\mathcal{U}(0,1)$ & $0.01891\substack{+0.00057\\-0.00052}$ & $\mathcal{U}(0,1)$ & $0.03082\substack{+0.00049\\-0.00050}$ \\
             b & & $\mathcal{U}(0,1)$ & $0.27\substack{+0.11\\-0.16}$ & $\mathcal{U}(0,1)$ & $0.056\substack{+0.053\\-0.037}$\\             
             K & (m/s) & $\mathcal{U}(0,100)$ & $2.71\substack{+0.66\\-0.66}$ & $\mathcal{U}(0,100)$ & $2.08\substack{+0.65\\-0.70}$\\
             \hline
             \multicolumn{6}{c}{Derived Parameters}\\
             \hline
             a/R$_*$ &  & - & $5.030\substack{+0.044\\-0.049}$ & - & $37.07\substack{+0.32\\-0.36}$\\
             $i$ & deg & - & \textcolor{black}{$86.9\substack{+1.3\\-1.8}$} & - & $89.914\substack{+0.083\\-0.057}$ \\
             R$_\text{P}$ & (R$_\oplus$) & - & $1.504\substack{+0.047\\-0.044}$ & - & $2.451\substack{+0.045\\-0.046}$\\
             a & (AU) & - & $0.01705\substack{+0.00022\\-0.00023}$ & - & $0.1257\substack{+0.0016\\-0.0017}$\\
             M$_\text{P}$ & (M$_\oplus$) & - & $3.49\substack{+0.85\\-0.85}$ & - & $7.27\substack{+2.27\\-2.45}$\\
             $\rho_\text{P}$ & (g/cm$^3$) & - & $5.64\substack{+1.48\\-1.46}$ & - & $2.71\substack{+0.86\\-0.93}$ \\
             T$_\text{eq}$ & (K) & - & 1478$\pm$20 & - & 544$\pm$7 \\
             S$_\text{P}$ & (S$_\oplus$) & - & $\textcolor{black}{791\pm43}$ & - & $14.57\substack{+0.79\\-0.80}$\\
             TSM & & - & $\textcolor{black}{43\pm11}$ & - & $\textcolor{black}{33\substack{+10\\-11}}$ \\
             \hline
             \hline
        \end{tabular}
        \label{tab:fit_results_planet}
\end{table*}
\subsection{Photometry}
\label{sec:photometry}
Around this star two planet candidates at periods of 1.05\,d and 42\,d were alerted by the \textit{TESS} team. To verify these, we perform our own search for the periods of the two planets using a Box Least Squares Periodogram \citep[BLS;][]{Kovacs_BLS_2002}. This analysis also found a planet at the same period as the \textit{TESS} team, $\sim$1.05\,d, with a log likelihood of logL=133. The second signal we picked up with logL=192 was at $\sim$21\,d as shown in green in \autoref{fig:bls}. This signal is half the period that was recorded by the \textit{TESS} vetting team.

A planetary signal of $\sim$21\,d is longer than the continuous \textit{TESS} observations which are downlinked every 13.7\,d. Therefore, $\sim$21\,d is a factor of 1.5 times the downlink and sector gaps. Hence, the even transits of the longer period sub-Neptune, TOI-2345\,c lie just at the edge of \textit{TESS}'s sector observation gaps and are only partially obtained. \textcolor{black}{We show these two even transits with the best fit model in \autoref{fig:tess_even_transits}}. This caused the \textit{TESS} vetting team to flag the system at twice the period. The \textit{CHEOPS} visit of the outer planet covers the even transit. Based on the transit times alone, we cannot differentiate between the 21\,d and 42\,d period. However, coupling the stellar density with the transit durations from \textit{TESS} and \textit{CHEOPS}, allows us to favour the 21\,d period. For this, we run a fit of the \textit{TESS} and \textit{CHEOPS} data in \textcolor{black}{\textsc{juliet}}  fitting for the period, P, mid-transit time, t$_0$, planet-to-star ratio, R$_\text{P}$/R$_*$, impact parameter, b and stellar density which are summarised in \autoref{tab:fit_results_planet}\textcolor{black}{.} Additionally we fit the jitter, the offset relative flux and the limb darkening coefficients parametrized following \citet{Kipping_Limb_Darkening_2013}, q$_1$ and q$_2$ for each photometric instrument, i.e. \textit{TESS} and \textit{CHEOPS} summarised in \autoref{tab:instr_fit}. We included a Gaussian Process (GP) with a Matern-3/2 kernel, implemented in \textcolor{black}{\textsc{juliet}}  \citep{Ambikasaran_george_2015,Foreman_Mackey_celerite_2017}, and fitted for the GP amplitude and time-scale as reported in \autoref{tab:instr_fit} to account for any residual systematic noise in the \textit{TESS} photometry. To determine the period of the outer planet we altered its period prior which we set to be uniformly distributed between 10\,d and 50\,d as this includes both possible periods. \textcolor{black}{Since the eccentricity can impact the transit duration, we also let it vary uniformly between 0 and 0.5 as well as the argument of periastron from 0° to 360° for the outer planet in this fit.}
As shown in \autoref{fig:tess}, the model picks up the signal of the outer transiting planet at $\sim$21\,d and fits the even transits which are only partially covered by \textit{TESS} observations. This model identifies the 21\,d planetary signal which also agrees with our previous BLS search. Hence, we determine a period of 21\,d for the outer planet \textcolor{black}{by analysing the TESS data carefully to spot transits close to the gaps and accounting for the stellar density in our fit. We use this period} for the further analysis. 

Using this period \textcolor{black}{and fixed eccentricity}, we run joint photometric analyses of the \textit{TESS} and \textit{CHEOPS} data \textcolor{black}{using priors as listed in} as listed in \autoref{tab:fit_results_planet}. We run two fits in order to compare the \textit{CHEOPS} aperture photometry obtained from the DRP and the \textit{CHEOPS} PSF photometry obtained by PIPE. In both analyses we apply the same GP as described above to the \textit{TESS} data. The fit using the \textit{TESS} and \textit{CHEOPS} DRP photometry results in radii of $1.48\pm0.05$\,R$_\oplus$ and $2.45\pm0.05$\,R$_\oplus$, while the \textit{TESS} and \textit{CHEOPS} PIPE photometry obtain radii of $1.48\substack{+0.04\\-0.05}$\,R$_\oplus$ and $2.48\pm0.06$\,R$_\oplus$ for the inner and outer planet respectively. These results are within a 1$\sigma$ agreement with each other. Since the PIPE photometry (median flux error=0.00052, RMS=0.00063) has a lower flux uncertainty and RMS than the DRP data (median flux error=0.00056, RMS=0.00089), we use the \textit{CHEOPS} PIPE data for the further analysis.

\begin{figure}
    \centering
    \includegraphics[width=\linewidth]{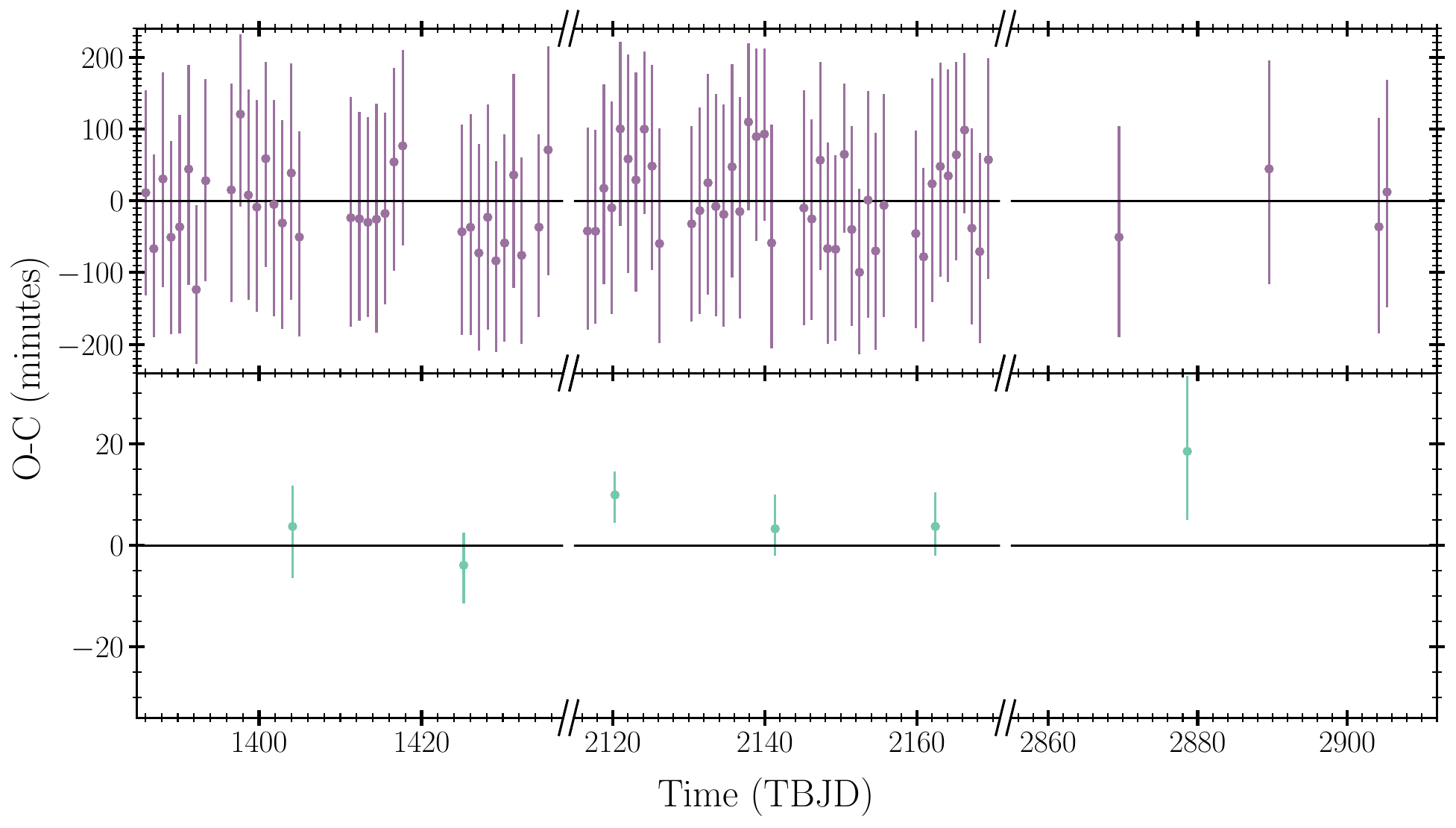}
    \caption{TTV Analysis of TOI-2345. Top: TOI-2345\,b in purple. Bottom: TOI-2345\,c in green.}
    \label{fig:ttvs}
\end{figure}

Within the \textit{TESS} and \textit{CHEOPS} photometric observations, we checked for transit timing variations that could be caused by potential further planets in the systems. For this we perform a further analysis on the \textit{TESS} and \textit{CHEOPS} data in \textcolor{black}{\textsc{juliet}}. In addition to fitting to the photometric priors used before, we also fit for a perturbation for each transit. This analysis does not identify any significant transit timing variations as shown in \autoref{fig:ttvs}. We note that the uncertainties on the transit times of the TOI-2345\,b are large. This is due to the very shallow transit, which is challenging for the transit model to identify in individual transits.

\subsection{Radial Velocity Time Series Assessment}
\label{subsec:Radial_Velocity_Fitting}

As described in \autoref{subsec:HARPS}, our \textcolor{black}{\textsc{s-bart}} re-extraction resulted in several sets of RVs extracted with different template matching and quality parameters. To determine the optimal RV time series for use in our joint fit, we fitted each of our RV sets in \textcolor{black}{\textsc{juliet}} and kept the period and mid-transit time fixed at the values obtained from the photometry only analysis given their precise determination. In our analysis we allow the semi-amplitudes and HARPS instrumentals (offset and jitter) to vary. Due to the number of RV data points being relatively low, we also kept the eccentricity and argument of periastron fixed at a circular orbit (e=0; $\omega$=90°). We selected the RV dataset that resulted in the lowest jitter in these fits. This was the case for the template matching parameters of an RV step of 0.5\, m/s and an RV limit of 200\,m/s using the \textcolor{black}{\textsc{s-bart}}'s classical method. This analysis selected \textcolor{black}{\textsc{s-bart}} outputs with quality checks of a minimum order SNR of 1.5, an airmass of 2.2 and RV errors of 5\,m/s. This set of RV data points has a median error of 2.04\,m/s and a RMS of 3.42\,m/s. \textcolor{black}{We show these re-extracted \textsc{s-bart} RVs in comparison to the DRS RVs in \autoref{fig:rv_timeseries}.}

\begin{figure}
    \centering    
    \includegraphics[width=\linewidth]{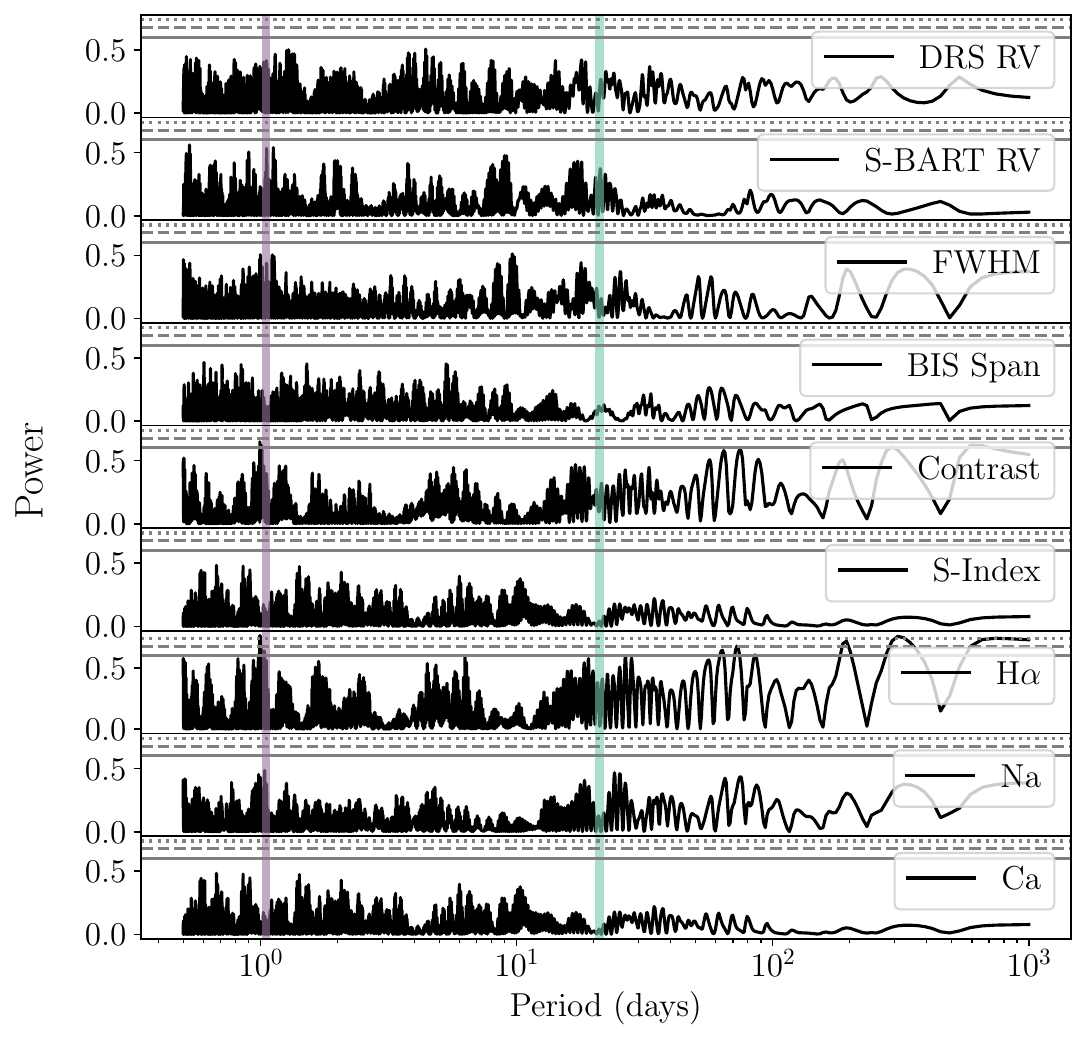}
    \caption{Lomb-Scargle periodograms of the \textcolor{black}{RVs (extracted from the HARPS DRS and \textsc{s-bart}) and} activity indicators recorded by the HARPS DRS. The false alarm levels of 1\%, 0.1\% and 0.01\% are shown by the gray straight, dashed and dotted line respectively. The orbital periods of the two planets are shown by the purple and green highlights.}
    \label{fig:activity_indicators}
\end{figure}

Additionally, we searched the activity indicators produced by the HARPS DRS for stellar activity signals using Lomb-Scargle periodograms \citep[][]{Lomb_1976,Scargle_1982} for each activity indicator. As in the photometric analysis of stellar activity, we do not find any significant peaks at the 1\% false alarm level or lower at typical rotation periods. As no activity indicators peak at the periods of the two planets (highlighted in purple and green in \autoref{fig:activity_indicators}), we conclude that the two signals in the RV data are indeed caused by the two planets and not stellar activity. \textcolor{black}{We also show Lomb-Scargle periodograms of the HARPS DRS and re-extracted \textsc{s-bart} RVs in \autoref{fig:activity_indicators} showing that there are no significant additional signals that could hint at RV only planets. Given the low number of RVs, there are no significant peaks at the orbital periods of TOI-2345\,b and TOI-2345\,c. } \textcolor{black}{Therefore, we conclude that it is fundamental to jointly fit the photometry and RVs to retrieve the masses for both planets.}

\subsection{Joint Fit}
We combine the photometric observations taken by \textit{TESS} and \textit{CHEOPS} with the radial velocity observations taken by HARPS and reprocessed using \textcolor{black}{\textsc{s-bart}} in a joint fit. As for our photometric analysis, we apply a GP as described in the photometry to the \textit{TESS} data. Since we did not find a stellar activity signal as shown in \autoref{subsec:rotation_period}, we do not use any GPs for the RV data. \\

Using \textcolor{black}{\textsc{juliet}}, we fit for the photometric priors as described in \autoref{sec:photometry} as well as the semi-amplitude, K, for planet b and c respectively which are summarised in \autoref{tab:fit_results_planet}. 

Additionally we fit the jitter the offset relative flux and the limb darkening coefficients parametrized following \citet{Kipping_Limb_Darkening_2013}, q$_1$ and q$_2$ for each photometric instrument, i.e. \textit{TESS} and \textit{CHEOPS}. For the HARPS, we fit the instrumental parameters of the offset and a jitter added in quadrature to the error bars summarised in \autoref{tab:instr_fit}. 

\textcolor{black}{Due to the limited number of RV observations, we run one fit with the eccentricity and the argument of periastron in our priors fixed resulting in a circular orbit for both planets (e=0, $\omega$=90°) and another fit where we let them vary uniformly between 0 and 0.5 as well as 0° and 360° respectively. The eccentric fit allows us to report a 3$\sigma$ upper limit on the eccentricity of 0.31. Since the circular fit results in a better lnZ (circular: lnZ=66320, eccentric: lnZ=66300), we report the results from the circular fit and use these values for the further analysis.}

\begin{figure*}
    \centering
    \includegraphics[width=0.8\linewidth]{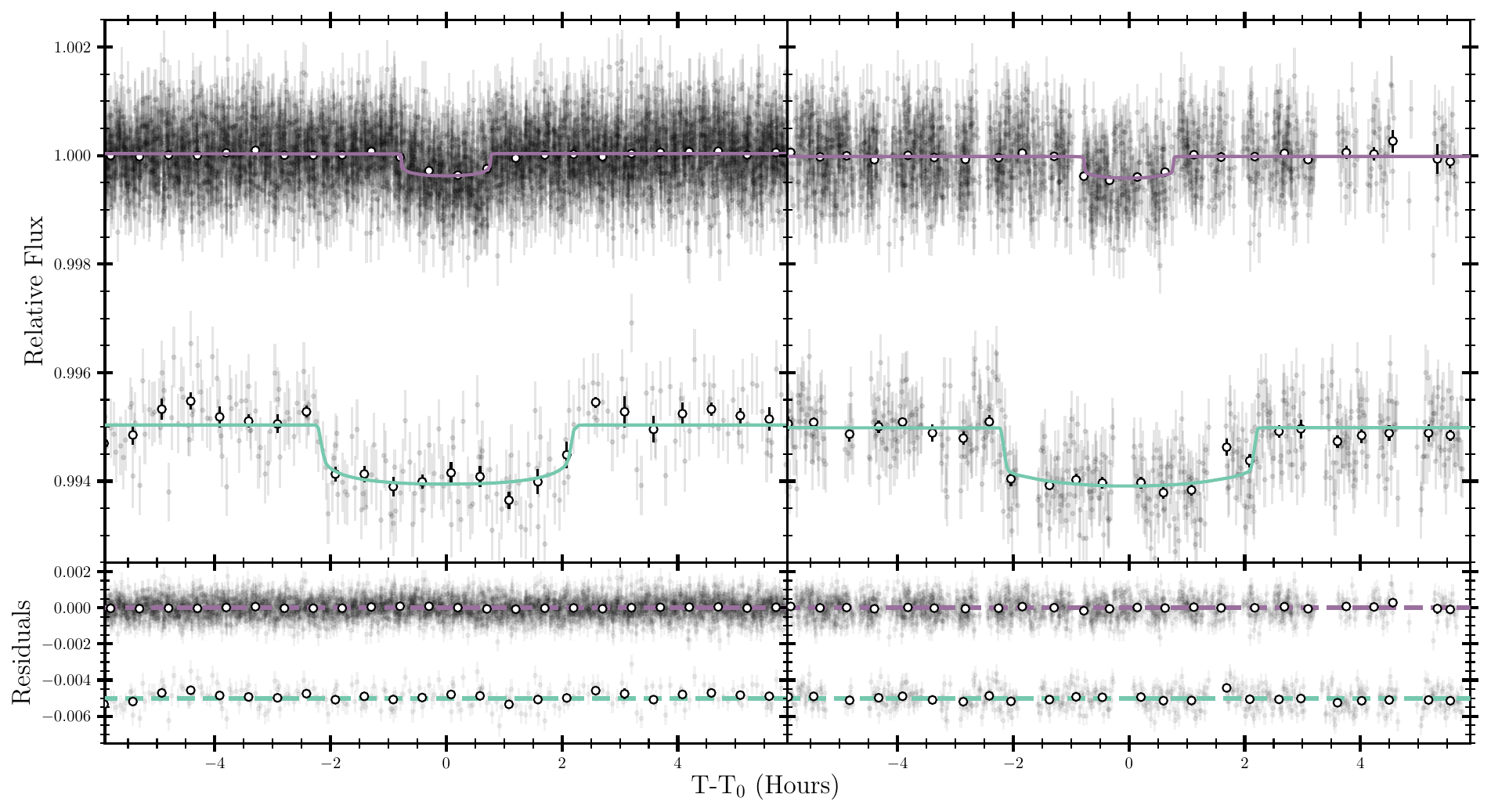}
    \caption{Phase folded transits of TOI-2345\,b (purple) and TOI-2345\,c (green) with the best-fit model from the joint fit in \textcolor{black}{\textsc{juliet}}  in the top panel. The bottom panel shows the residuals. Left: \textit{TESS} data. Right: \textit{CHEOPS} data.}
    \label{fig:transits}
\end{figure*}

\begin{figure*}
    \centering    \includegraphics[width=0.8\linewidth]{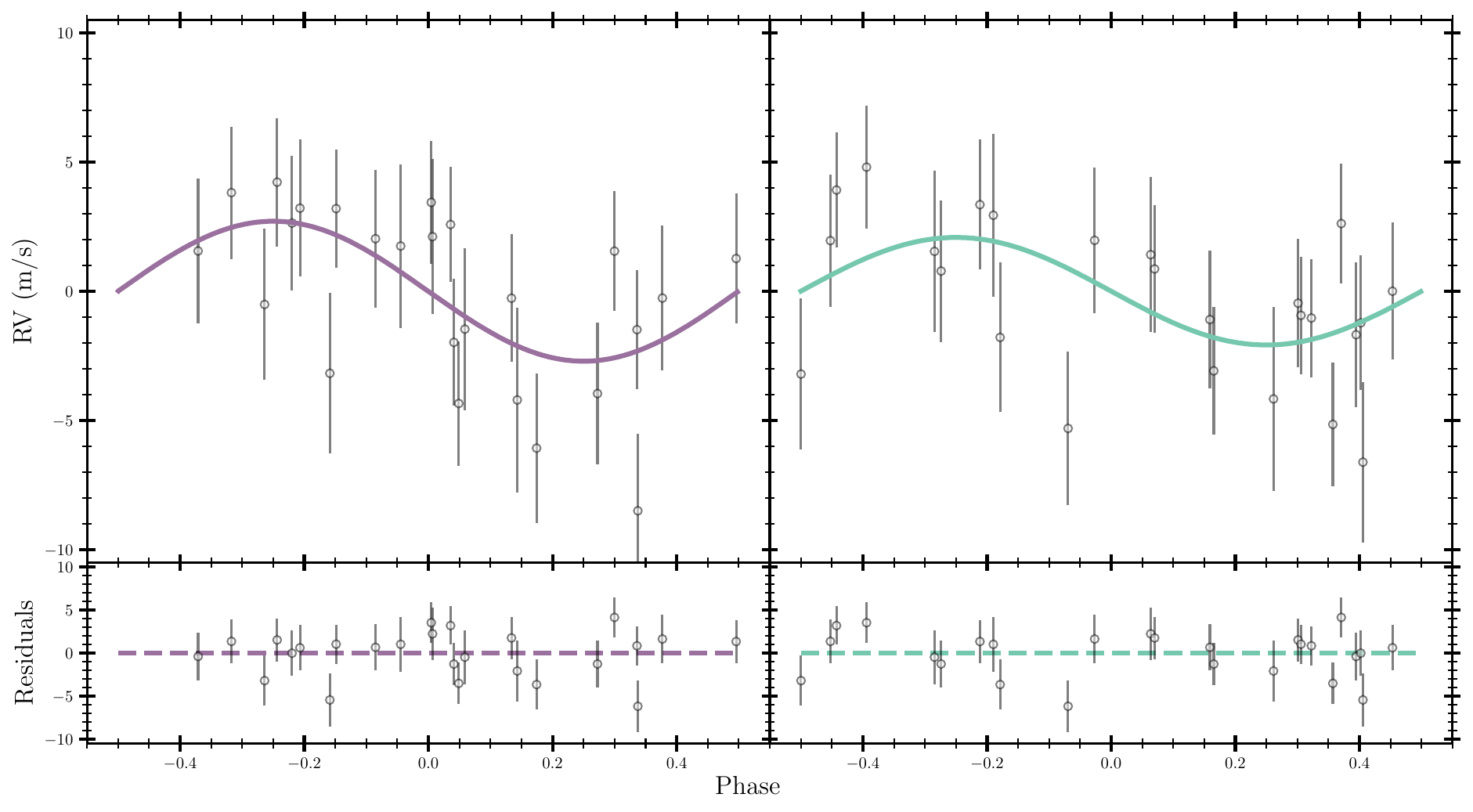}
    \caption{Phase folded RV curves of the HARPS data re-extracted using \textcolor{black}{\textsc{s-bart}} of TOI-2345\,b in purple on the left and TOI-2345\,c in green on the right. The best fitted model from the joint fit in \textcolor{black}{\textsc{juliet}} in the top panel. The bottom panel shows the residuals.}
    \label{fig:rvs}
\end{figure*}

We obtain a radius of $1.504\substack{+0.047\\-0.044}$\,R$_\oplus$ and mass of $3.49\pm0.85$\,M$_\oplus$ for the inner planet, TOI-2345\,b, and a radius of $2.451\substack{+0.045\\-0.046}$\,R$_\oplus$ and mass of $7.27\substack{+2.27\\-2.45}$\,M$_\oplus$ for the outer planet, TOI-2345\,c. The fitted and derived values are summarised in\autoref{tab:fit_results_planet}. The phase-folded transits and RV curves of this joint fit are shown in \autoref{fig:transits} and \autoref{fig:rvs} respectively. Given the large upper bounds on the GP hyper-parameters, we conclude that the GP applied to the \textit{TESS} data accounts for residual noise in the photometry.

\section{Discussion}
\label{sec:Discussion}

\subsection{Interior Structure Modelling}

\begin{figure*}
    \centering
    \includegraphics[width=0.9\linewidth]{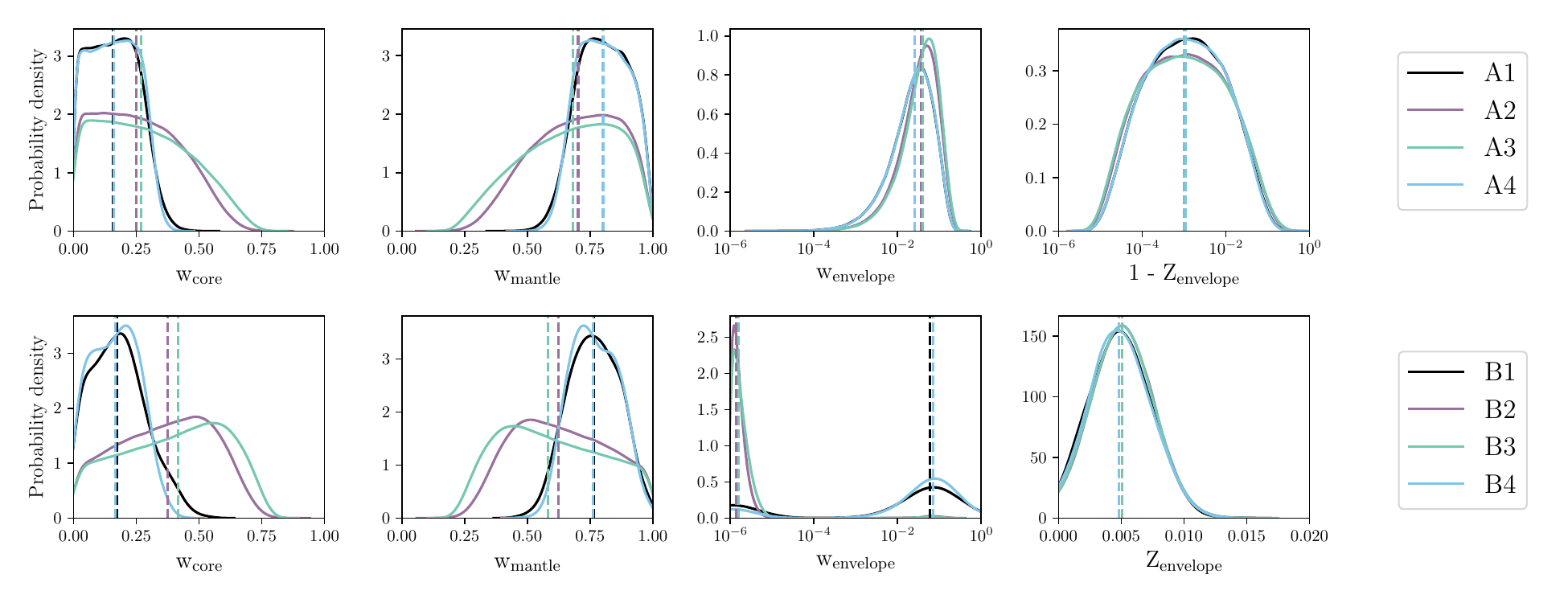}
    \caption{plaNETic posterior distributions for TOI-2345\,b. The different abundance priors are shown by the different colours.  The median of each posterior is marked by the vertical dashed line. Top: Water-rich prior. Bottom: Water-poor prior.}
    \label{fig:plaNETic_TOI2345b}
\end{figure*}

\begin{figure*}
    \centering
    \includegraphics[width=0.9\linewidth]{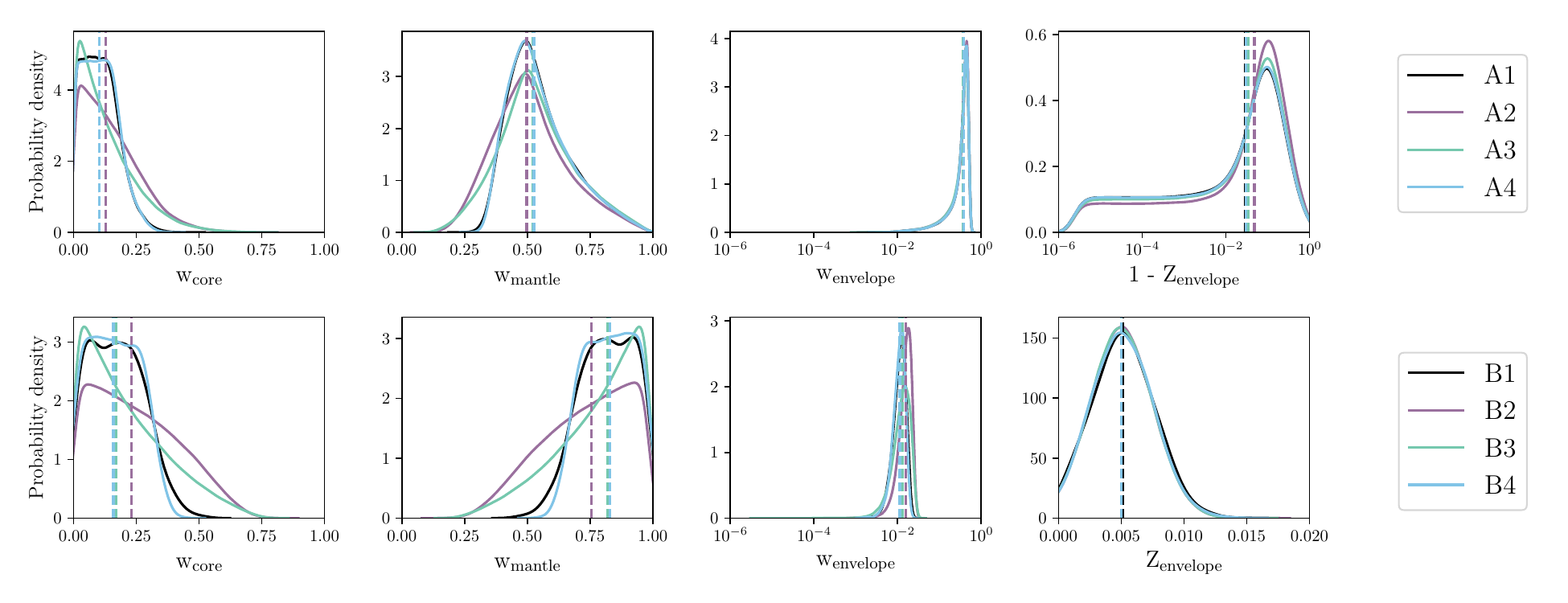}
    \caption{plaNETic posterior distributions for TOI-2345\,c. The different abundance priors are shown by the different colours.  The median of each posterior is marked by the vertical dashed line. Top: Water-rich prior. Bottom: Water-poor prior.}
    \label{fig:plaNETic_TOI2345c}
\end{figure*}

Using the stellar and planetary parameters of the system, we model the interior structure of the two planets using plaNETic \citep{Egger_TOI469_plaNETic_2024}. plaNETic is an interior structure modelling framework that was first introduced in the analysis of the three planets orbiting TOI-469 \citep[][]{Egger_TOI469_plaNETic_2024}. As these planets are also in the super-Earth and sub-Neptune regime, this interior structure modelling is tailored well to our planets. plaNETic is using Deep Neural Networks which are trained on mass regimes from 0.5-6\,M$_\oplus$ and 6-15\,M$_\oplus$ of which both of the planets around TOI-2345 fall. Additionally, each mass range has two different interior structure model databases, one with a water rich and one with a water poor prior represented by the top and bottom row in \autoref{fig:plaNETic_TOI2345b} and \autoref{fig:plaNETic_TOI2345c} respectively.

Within the high and low water prior models, three different priors for the planetary abundances of Fe, Mg and Si are modelled. These vary from being the same as the stellar abundances as suggested by \citet{Thiabaud_elements_stars_planets_2015}, iron-enriched as suggested by \citet{Adibekyan_compositional_link_2021} or uniformly sampled with an upper limit of 75\% of Fe compared to the other two refractory elements. 

As stars and planets are formed from the same material \citep{Nielsen_Planet_Formation_2023} there is a strong connection between the abundances of the host star and the planets which can also be seen for the Sun and the Earth \citep{Wang_Protosun_Abundances_2019}. In plaNETic, this is accounted for by the first composition prior. However, this is only the case for refractory elements and not volatiles. Indeed, \citet{Adibekyan_compositional_link_2021} showed that the connection between stellar and planetary composition may not be one-to-one. \citet{Wang_ExoInt_2019} developed a tool, ExoInt, to devolatilise stars and obtain the abundances of elements in orbiting planets. Hence we apply ExoInt to our host star in order to obtain the devolatised element abundances that are expected to make up the bulk composition of its orbiting planets. We use the stellar abundances for [Fe/H], [Si/H] and [Mg/H] as recorded in \autoref{tab:stellarParam}. ExoInt also requires an abundance of [O/H]. As TOI-2345 is a cool star, it is challenging to determine the oxygen abundance well from the obtained HARPS spectra. So we use the public APOGEE DR17 \citep{Niveder_APOGEE_Data_Reduction_2015,Majewski_APOGEE_Overview_2017,Abdurrouf_APOGEE_DR17_Release_2022} catalogue and select stars with the TOI-2345 $T_{\mathrm{eff}}$, $\log{g}$ and [Fe/H] values within the uncertainties. We also remove stars of bad quality flags as recommended by APOGEE. This leaves us with 1265 out of the $\sim$700000 stars with recorded abundances in APOGEE DR17. Averaging these and their uncertainties, we obtain [O/H]=-0.09$\pm$0.07. Using these stellar abundances, we devolatise TOI-2345 using ExoInt. This results in new abundance ratios for the refractory elements of the planets that may reflect the devolatised abundance ratios in the protoplanetary disk better than assuming purely stellar abundances. ExoInt and plaNETic both record the abundances as 10$^{[{\rm X/Fe}]_*}$, where X is Si or Mg. For the stellar abundances in plaNETic, we obtain these to be 1.12 for Si and 1.62 for Mg; using ExoInt they are 1.00 and 1.63 for Si and Mg respectively. We use these planetary abundances for Mg and Si from ExoInt as an additional prior in plaNETic. 

We find the result of this additional model as shown by the pink line in \autoref{fig:plaNETic_TOI2345b} and \autoref{fig:plaNETic_TOI2345c} to be in agreement with the other three priors of plaNETic. Especially, we note that the ExoInt priors (A4 and B4) are similar to the stellar abundances (A1 and B1). However they are slightly higher in the core and lower in the mantle which is due to the devolatising of ExoInt and hence having slightly higher abundances of refractory (heavier) elements in our priors. As summarised in \autoref{tab:planetic_results}, we find from our interior structure analysis that the inner planet has a low core and high mantle mass fraction for the water-poor and water-rich priors. As expected for a highly irradiated super-Earth its atmospheric mass fraction is very low and nominally consistent with 0 for both sets of priors. The outer planet has a similar distribution of core and mantle mass to the inner planet in the water-poor prior. However, the water-rich prior increases the water content of this planet significantly while shrinking the core and mantle mass fractions. Since we do not know the formation path of these planets, we cannot differentiate between the water-rich and water-poor priors. 

\begin{table*}
    \caption{Core, Mantle, Water and Atmospheric Mass Fraction modelled by plaNETic for planet b and c using devolatised stellar abundances from ExoInt (4) and the water-rich (A) and water poor (B) priors.}
    \centering
    \begin{tabular}{cccccc}
    \hline
    \hline
         Planet & Model & CMF & MMF & WMF & AMF \\
         \hline
         TOI-2345\,b & A4 & $0.16\substack{+0.11\\-0.11}$ & $0.80\substack{+0.11\\-0.11}$ & $0.026\substack{+0.043\\-0.019}$ & $0.000021\substack{+0.000185\\-0.000018}$ \\
         TOI-2345\,b &  B4 & $0.167\substack{+0.099\\-0.111}$ & $0.763\substack{+0.107\\-0.094}$ & $0.00028\substack{+0.00034\\-0.00028}$ & $0.070\substack{+0.039\\-0.070}$\\
         TOI-2345\,c & A4 & $0.10\substack{+0.074\\-0.071}$ & $0.526\substack{+0.158\\-0.098}$ & $0.367\substack{+0.094\\-0.213}$ & $0.011\substack{+0.019\\-0.011}$ \\
         TOI-2345\,c & B4 & $0.16\substack{+0.12\\-0.11}$ & $0.83\substack{+0.11\\-0.11}$ & $0.000053 \substack{+0.000038\\-0.000029}$ & $0.0113\substack{+0.0034\\-0.0036}$ \\
    \hline
    \hline
    \end{tabular}
    
    \label{tab:planetic_results}
\end{table*}

\subsection{Atmospheric Evolution Modelling}

We evaluated the atmospheric evolution of TOI-2345\,b and c using the \textsl{P}lanetary \textsl{A}tmosphere and \textsl{S}tellar Ro\textsl{T}ation R\textsl{A}te \citep[\textsc{PASTA};][]{bonfanti2021b} code. The tool works in the Markov chain Monte Carlo (MCMC) framework \textsc{MC3} developed by \citet{cubillos2017} and at each chain step and for each exoplanet it computes the evolutionary track of the atmospheric mass fraction $f_{\mathrm{atm}}(t)$ from the time of dispersal of the proto-planetary disk $t_{\mathrm{disk}}$, assumed to be 5 Myr, up to the present age of the system $t_{\star}$. To do so, \textsc{PASTA} assumes that: (i) no migration has occurred after $t_{\mathrm{disk}}$, (ii) the atmosphere is H-dominated, and (iii) the atmospheric mass loss is driven by thermal mechanisms, that is high energy (X-ray + EUV: XUV) flux from the stellar host \citep[photo-evaporation; see e.g.][]{lammer2003,chen2016} and core-powered mass loss \citep[e.g.][]{Ginzburg_Core_Powered_Mass_Loss_2018,Gupta_Schlichting_Core_Powered_Mass_Loss_2019}.

\textsc{PASTA} is model-dependent and uses the MESA isochrones and stellar tracks \citep[MIST;][]{choi2016} to trace the evolution of stellar parameters over time, a gyrochronological relation $P_{\mathrm{rot}}(t)$ in the form of a broken-power law as presented in \citet{bonfanti2021b}, a set of empirical relations to convert the stellar rotation period $P_{\mathrm{rot}}$ into the emitted XUV-flux at any given epoch \citep[][and references therein]{bonfanti2021b}, a model for computing the atmospheric mass loss \citep{kubyshkina2018aMdot,kubyshkina2018bMdot}, and a planetary structure model that links the planetary observables with the expected atmospheric content \citep{johnstone2015planetModels}. \textsc{PASTA} requires input parameters both on the stellar side ($M_{\star}$, $t_{\star}$, and $P_{\mathrm{rot},\star}$) and on the planetary side ($M_p$, $R_p$, and $a$ for each exoplanet in the system). The present-day stellar rotation period $P_{\mathrm{rot},\star}$ has been inferred by inverting the gyrochronological relation from \citet{Barnes_Kim_Stellar_Activity_2010}, while all other system parameters are taken from Table~\ref{tab:stellarParam} and \ref{tab:fit_results_planet}. 
Once a given MCMC step has been performed, for each planet the code builds its $f_{\mathrm{atm}}(t)$, it converts it into the corresponding evolutionary track in radius $R_p(t)$ according to the planetary structure models and checks whether $R_p(t_{\star})$ is consistent with the observed $R_p$. If so, that chain step is accepted and the consequent $f_{\mathrm{atm}}(t_{\mathrm{disk}})\equiv f_{\mathrm{atm}}^{\mathrm{start}}$ value builds up the posterior distribution for the initial atmospheric mass fraction of the planet. As a by-product, we also get the stellar rotation period when the star was young. 

\begin{figure*}
\centering
\includegraphics[width=\textwidth]{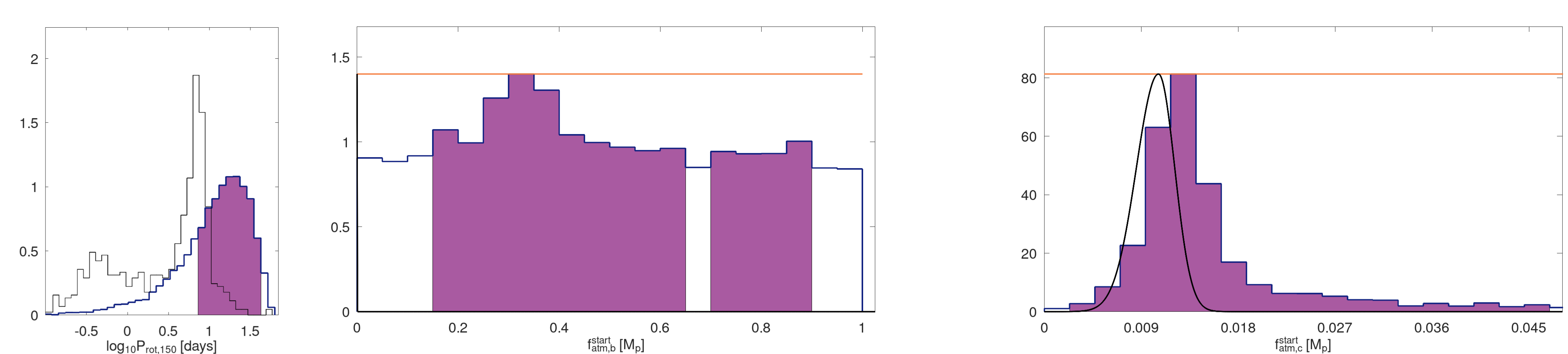}
\caption{Posterior distributions (thick blue histograms with the underneath area normalised to 1) as outputted by \textsc{PASTA} for the stellar rotation period when the star was 150 Myr old $P_{\mathrm{rot},150}$ (\textit{left panel}) and for the initial atmospheric mass fraction $f_{\mathrm{atm}}^{\mathrm{start}}$ of TOI-2345\,b (\textit{middle panel}) and TOI-2345\,c (\textit{right panel}). The purple area represent the 68\% highest probability density intervals. The thin black histogram in the left panel depicts the $P_{\mathrm{rot},150}$ distribution of open-cluster stars with masses comparable to that of TOI-2345 as taken from \citet{johnstone2015Prot150}. The black lines in the \textit{middle} and \textit{right} panels represent the present-day atmospheric content of the planets (which is negligible for TOI-2345\,b), while the orange horizontal lines mark that uniform priors were imposed on the $f_{\mathrm{atm}}^{\mathrm{start}}$ jump parameters. This is in agreement with our interior structure modelling results from plaNETic which also found the inner planet to have a negligible atmosphere, while the atmospheric mass fraction of the outer planet agrees with the present day value found from the atmospheric evolution modelling. }
\label{fig:Prot_fatm}
\end{figure*}

Fig.~\ref{fig:Prot_fatm} (left panel) compares the posterior distribution of $P_{\mathrm{rot}}$ when the star was 150 Myr old (thick blue histogram) with the $P_{\mathrm{rot}}$ distribution of stars belonging to 150-Myr-old open clusters and having a mass comparable to $M_{\star}$ \citep[thin black histogram; taken from][]{johnstone2015Prot150}, which suggests that TOI-2345 was likely born as a slow rotator. The other two panels of Fig.~\ref{fig:Prot_fatm}, instead, show the posterior distribution of $f_{\mathrm{atm}}^{\mathrm{start}}$ for planet b and c, in comparison with the present-day atmospheric content expected if the atmosphere is H-dominated (black line). In the middle panel, the black line is basically compatible with zero (i.e. no atmosphere is expected around the highly irradiated USP) and this condition is reached regardless of $f_{\mathrm{atm}}^{\mathrm{start}}$. Finally, the right panel shows that the present-day $f_{\mathrm{atm}}$ of TOI-2345\,c is similar to the inferred $f_{\mathrm{atm}}^{\mathrm{start}}$, with the planet that has likely lost only a small fraction of its initial atmospheric content.

\subsection{Assessment with Synthetic Planet Population}

To understand the broader formation and evolution conditions that the TOI-2345 planets have undergone, we placed the system in the context of a synthetic planet population. As host star properties have a strong impact on the underlying physical processes that impact orbiting planets, we retrieved the results of a synthetic population that most closely match the spectral type of TOI-2345 \citep{Burn2021}. These simulations of 999 K-dwarf systems each with 50 initial planetary embryos were produced by a detailed formation and evolution framework \citep{Emsenhuber_Bern_model_2021}. In brief, the embryos undergo formation and growth in a coupled gas and dust disc with the amount of refractory and volatile elements accreted during formation being monitored. This is useful in understanding the interior structures of these synthetic planets. Within the theoretical framework evolution is handled via atmospheric escape, giant impacts, and gas-driven, tidal-forces, and planet-planet scattering migration. Importantly, the evolution of the synthetic planetary systems is also coupled to the evolution of the host star from formation to the end state of the simulation at 5\,Gyr.

For our comparative analyses, we selected TOI-2345-like systems following the procedures outlined in \citet{Egger_TOI469_plaNETic_2024}. This was done by taking the TOI-2345\,b and c radii, masses, and semi-major axes derived from our joint fit (see Table~\ref{tab:fit_results_planet}), and choosing synthetic planets within 25\% or 3$\sigma$ (choosing the larger) of the TOI-2345 planets.

To test the tentative stellar-planetary compositional link, we assessed the interior structures of the synthetic planets as a function of stellar metallicity. To do so, we further restrict the TOI-2345\,b and c-like planets to those that orbit a star with a similar [Fe/H] to that of TOI-2345. This results in eight TOI-2345\,b-like bodies and 16 TOI-2345\,c-like planets. For the inner planets, we find core-to-mantle mass ratios of 0.24-0.32 which agrees within 1$\sigma$ of our value (0.20$\pm$0.14) from our A4 interior structure modelling reported in \autoref{tab:planetic_results}. Interestingly, the TOI-2345\,c-like planets that orbit a compositionally-similar star have more constrained core-to-mantle mass ratios of 0.24-0.25 with a wider range of water and atmospheric mass fractions. Whilst again slightly lower than predicted values, our A4 interior structure modelling core-to-mantle mass ratio for TOI-2345\,c (0.19$\pm$0.15) also agrees with the synthetic planets.

Given the wide period gap between the two detected TOI-2345 planets, we aimed to search the synthetic systems to understand the rarity of such a configuration when considering the formation and evolution processes include in the framework \citep{Emsenhuber_Bern_model_2021}. To do this, we selected multi-planet synthetic systems that host at least two bodies that resemble TOI-2345\,b and c without the stellar metallicity constraint. Of the 999 systems, only one appears TOI-2345-like via this criteria. However, this synthetic system also includes two additional planets between the TOI-2345\,b and c-like bodies with orbital periods of $\sim$4.0 and 11.5\,d. The masses of these bodies ($\sim$7.1 and 11.5\,M$_\oplus$) result in RV semi-amplitudes of $\sim$4.6 and 5.4\,m/s, respectively. Should such an architecture be present in the TOI-2345 system it would be detectable in our HARPS data as the RMS scatter of the residuals to our HARPS fit is $\sim$2.6\,m/s. Given the simulated co-planarity, these additional planets would also be apparent in the \textit{TESS} photometry. However, no such signals exist. While the number of simulated systems we compared to TOI-2345 is relatively small and the theoretical framework may not include all important processes (such as pebble accretion or MHD wind driven disks), it could still provide a good population overview. Therefore, the TOI-2345 system may be considered architecturally rare via this comparison and worthy of future dynamical studies to understand the migration history.

\subsection{TOI-2345 in the context of thick disk systems}

\begin{figure*}
    \centering
    \includegraphics[width=1\linewidth]{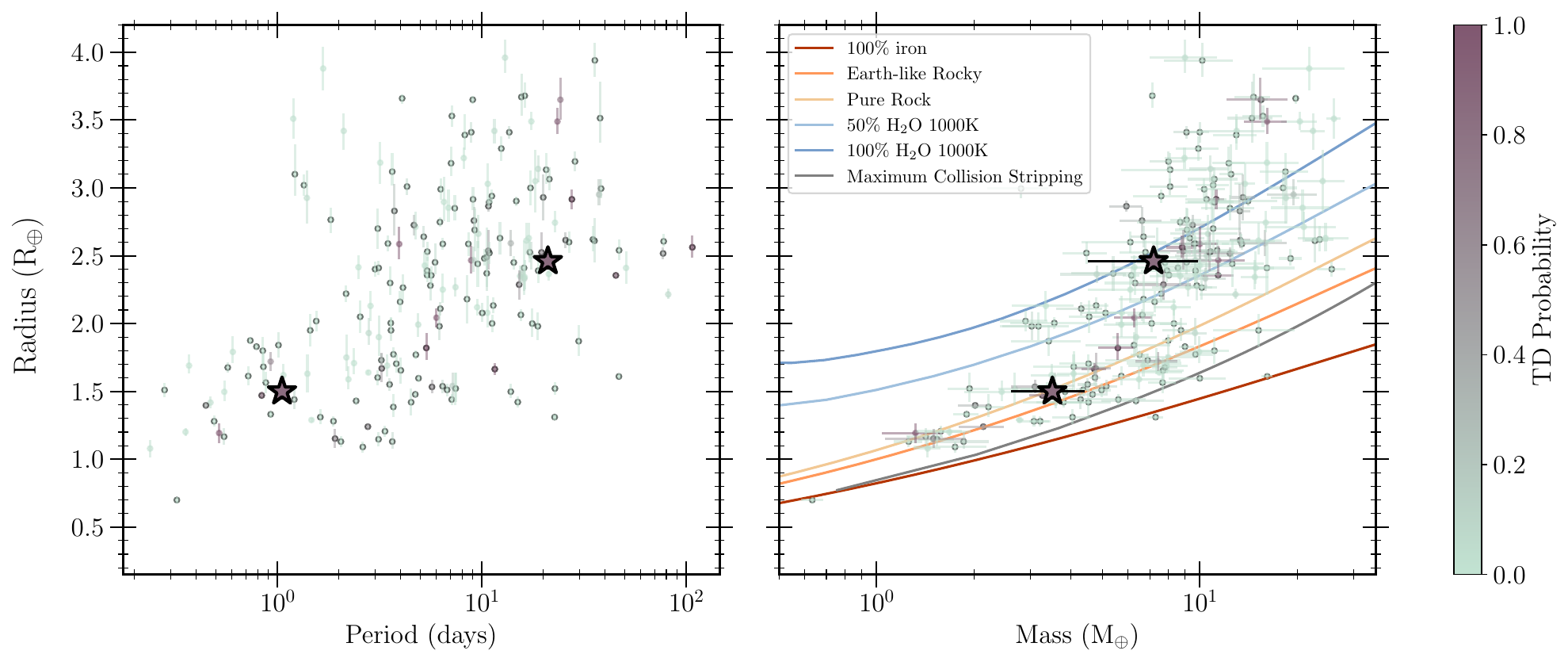}
    \caption{Comparison of the two planets orbiting TOI-2345 to other well characterised planets below 4\,R$_\oplus$ colour-coded by the respective weighted kinematic thick disk probability of the host star. TOI-2345 is shown by the stars in purple with the black border, where the colour is representing its high thick disk probability. Multi-planetary systems are highlighted by circles. Left: Radius-Period diagram. Right: Radius-mass diagram. We show compositional lines for 100\% iron, Earth-like rocky, pure rock, 100\% water at 1000\,K and 50\% water at 1000\,K at the respected colour-coded lines following \citet{Zeng_Growth_Model_2019}. Additionally, we show the maximum collision stripping following \citet{Marcus_Giant_Impacts_Super_Earths_2010} by the gray line. }
    \label{fig:td_comparison}
\end{figure*}

As computed in \autoref{sec:toi2345_kinematics}, TOI-2345 is kinematically a thick disk star (P(TD)=85\%). In order to compare this system to other kinematic thick disk systems, we download all known exoplanets from the NASA Exoplanet Archive \citep{Exoplanet_Archive}. As we rely on the stellar kinematic properties recorded by \textit{Gaia} to compute kinematic probabilities, we also filter for targets that have a recorded \textit{Gaia} ID, which leaves us with 5605 planets. To focus on systems hosting small planets that are well characterised, we apply cut-offs of V<13\,mag and R$_\text{P}$<4\,R$_\oplus$ determined with a precision better than 10\%. To make sure we only include systems with a measured mass we filter for the mass provenance flag to be "mass". We only include systems with M$_\text{P}$<30\,M$_\oplus$ and measured with a precision better than 30\%, similar to our inner planet. This leaves us with 207 planets. In our stellar kinematic analysis we use the most recent \textit{Gaia} DR3 data, however as the NASA Exoplanet Archive used \textit{Gaia} DR2, we cross-match the \textit{Gaia} DR2 IDs with the \textit{Gaia} DR3 IDs using TOPCAT \citep{Taylor_TOPCAT_2005} and \textit{Gaia}-ARI \citep{Gaia_DR3_Release_Summary_2023}. Following the process described in \autoref{sec:toi2345_kinematics}, we compute the kinematic probabilities for each of these systems \textcolor{black}{which are shown in \autoref{fig:td_comparison}}. 

Out of the remaining 207 planets, we find that 19 planets are orbiting a host star with a kinematic thick disk probability of above 50\%. 12 of these are in multiplanetary system around 6 different host stars. In one of these systems, K2-111 \citep{Mortier_K2_111_2020}, the second planet is not transiting and consequently does not have a radius measurement. The second planet orbiting HIP 9618 is only detected with a 3-$\sigma$ upper mass limit \citep{Osborn_HIP9618_2023}. Hence we do not include these two systems in our comparison. The remaining 4 well-characterised systems are HD 136352 with a computed weighted thick disk probability of 92\% orbited by three small planets \citep{Udry_HARPS_XLIV_2019,Delrez_HD136352_2021}, Kepler-10 with a thick disk probability of 79\% that hosts two transiting and one non-transiting small planets \citep{Batalha_Kepler10b_2011,Fressin_Kepler10c_2011,Bonomo_Kepler10_2025}, TOI-178 that has a thick disk probability of 55\% and 6 small planets out of which 3 are well characterised to fulfil our comparison criteria \citep{Leleu_TOI178_2021}, and HIP 8152 with a computed weighted thick disk probability of 52\% orbited by 2 small planets \citep{Akana_Murphy_TESS_Keck_2023,Polanski_TESS_Keck_2024}. As there are only 4 thick disk systems that contain two or more well-characterised transiting planets below 4\,R$_\oplus$, the discovery of the two planets transiting TOI-2345 adds to this limited sample. Among these systems it has the second highest thick disk probability, just below HD 136352, and falls within the range of stellar metallicities.

\begin{figure}
    \centering
    \includegraphics[width=\linewidth]{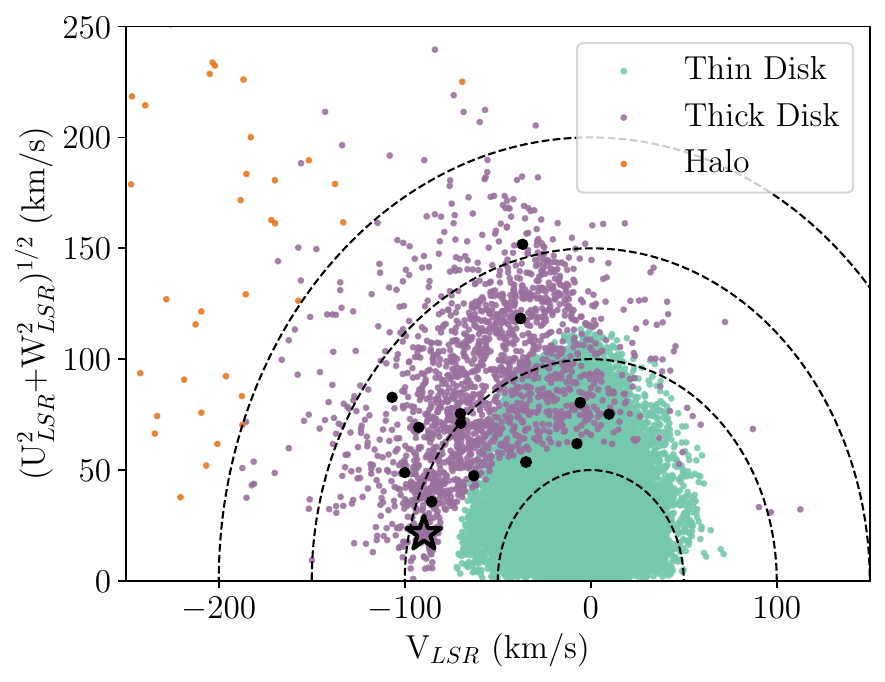}
    \caption{Toomre Diagram of the stellar sample and galactic classifications of \citet{Chen_PAST_II_2021}. Stars of the thin disk are shown in green, stars of the thick disk in purple and stars of the halo in orange. The well-characterised thick disk systems from \autoref{tab:thick_disk_planets} are shown in black of which the kinematics are computed using the LSR from \citet{Tian_Stellar_Kinematic_LAMOST_LSR_2015}. TOI-2345 is shown by the purple star with the black border and clearly placed in the thick disk.}
    \label{fig:toomre}
\end{figure}

We demonstrate the location of TOI-2345 in a Toomre diagram in \autoref{fig:toomre} in which we plot the kinematic properties and galactic classifications of the LAMOST-\textit{Gaia}-\textit{Kepler} sample from \citet{Chen_PAST_II_2021}. Placing TOI-2345 in context of the other well-characterised systems from \autoref{tab:thick_disk_planets} which are shown in black, we find that TOI-2345 is clearly kinematically a thick disk star. 

In order to understand the population of small planets around thick disk stars further and explore trends in their composition, the sample has to be expanded and TOI-2345 contributes to this. For example, the well-characterised interior structures of the TOI-2345 planets can be used to test planet density predictions for thick disks stars in galactic chemical evolution models \citep{Steffen_Galactic_Chemical_Evolution_2025}, and assess planet frequency statistics across stellar kinematic and chemical families \citep{Nielsen_Planet_Formation_2023}. \citet{Kruijssen_Phase_Space_2020} predicted that planets below the radius valley should be less frequent around low-density phase space stars that are predominantly found in the Galactic thick disk \citep{Mustill_Phase_Space_2022}. The super-Earth and sub-Neptune orbiting TOI-2345 span the radius valley \textcolor{black}{as shown in \autoref{fig:td_comparison}} and so provide evidence to refute this prediction. This is also found for HD 136352, Kepler-10, and TOI-178. However, also this is still a very restricted sample and so importantly TOI-2345 adds to this. 

Additionally, TOI-2345\,b is on a 1\,d orbit and hence also increases the sample of USPs around thick disk stars. In combination with the wider orbital spacing of planets\,b and c, the TOI-2345 system is in agreement with the broader stellar kinematic-planet architecture trends seen in a \textit{Kepler}-based sample of USPs \citep{Winn_Kepler78_USP_2018,Tu_USP_Age_Architecture_2025}. Following \citet{Castro_Gonzales_K2_157_2025}, USPs could have formed through high-eccentricity migration \citep{Petrovich_USP_high_eccentricity_2019}, low-eccentricity migration \citep{Pu_Lai_USP_low_eccentricity_2019} or obliquity-driven migration \citep{Millholland_USP_Obliquity_2020}. As we do not find a high eccentricity for TOI-2345\,b which could hint at high-eccentricity migration, nor planetary neighbours in close-in orbits in <10\,d, which are expected in a obliquity-driven migration scenario, the most likely formation of TOI-2345 is the low-eccentricity scenario.

\begin{table*}
    \begin{center}
    \begin{tabular}{lllllllll}
    \hline
    \hline
       Planet  &  R$_\text{P}$ & M$_\text{P}$ & $\rho_\text{P}$ & T$_\text{eq}$ & P(TD) & [Fe/H] & \# planets & Reference \\
        &  (R$_\oplus$) & (M$_\oplus$) & (g/cm$^3$) & K &  &  &  &  \\
       \hline
       K2-111\,b & $1.820\substack{+0.11\\-0.090}$ & $5.58\substack{+0.74\\-0.73}$ & $5.0\substack{+1.1\\-1.0}$ & $1309\pm19$ & 0.991 & -0.46$\pm$0.05 & 2 & [1,2] \\
       GJ 1252\,b & $1.193\pm0.074$ & $1.32\pm0.28$ & $4.2\substack{+1.3\\-1.1}$ & $1089\pm69$ & 0.987 & +0.1$\pm$0.1 & 1 & [3,4] \\
       TOI-669\,b & $2.59\substack{+0.13\\-0.11}$ & $10.0\pm1.4$ & $3.17\substack{+0.73\\-0.95}$& $1125\pm19$ &  0.965 & -0.06$\pm$0.09 & 1 & [5,6] \\
       Wolf 503\,b & $2.043\pm0.069$ & $6.27\substack{+0.85\\-0.84}$ & $4.03\substack{+0.72\\-0.64}$ &  $789\pm16$ & 0.965 & -0.47$\pm$0.08 & 1 & [7,2] \\
       K2-180\,b & $2.466\substack{+0.110\\-0.096}$ & $11.4\substack{+2.4\\-2.2}$ & $4.16\substack{+1.10\\-0.92}$ & $797.5\substack{+8.4\\-8.5}$ & 0.953 & -0.65$\pm$0.10 & 1 & [8,9] \\
       HD 136352\,b & $1.664\pm0.043$ & $4.72\pm0.42$ & $5.62\substack{+0.72\\-0.66}$ & $905\pm14$ & 0.926 & -0.24$\pm$0.05 & 3 & [10,11] \\
       HD 136352\,c & $2.916\substack{+0.075\\-0.073}$ & $11.24\substack{+0.65\\-0.63}$ & $2.50\substack{+0.25\\-0.23}$ & $677\pm11$ & 0.926 & -0.24$\pm$0.05 & 3 & [10,11] \\
       HD 136352\,d & $2.562\substack{+0.088\\-0.079}$ & $8.82\substack{+0.93\\-0.92}$ & $2.88\substack{+0.43\\-0.40}$ & $431\pm7$ & 0.926 & -0.24$\pm$0.05 & 3 & [10,11] \\
       TOI-5076\,b & $3.489\substack{+0.100\\-0.094}$ & $16.1\pm2.5$ & $2.08\substack{+0.35\\-0.34}$ & $550\pm14$ & 0.921 & +0.20$\pm$0.08 & 1 & [12,13] \\
       \hline
       TOI-2345\,b & $1.504\substack{+0.047\\-0.044}$ &  $3.49\pm0.85$ & $5.64\substack{+1.48\\-1.46}$ & 1478$\pm$20 & 0.850 & -0.10$\pm$0.07 & 2 & This work \\
       TOI-2345\,c & $2.451\substack{+0.045\\-0.046}$ & $7.27\substack{+2.27\\-2.45}$ & $2.71\substack{+0.86\\-0.93}$ & 544$\pm$7 & 0.850 & -0.10$\pm$0.07 & 2 & This work\\
       \hline
       Kepler-10\,b & $1.47\substack{+0.03\\-0.02}$ & $3.24\pm0.32$ & $5.54\substack{+0.66\\-0.62}$ & $2188\pm16$ & 0.789 & -0.15$\pm$0.04 & 3 & [14,15] \\
       Kepler-10\,c & $2.355\pm0.022$ & $11.29\pm1.24$ & $4.75\pm0.53$ & $579\pm4$ & 0.789 & -0.15$\pm$0.04 & 3 & [16,15] \\
       TOI-1231\,b & $3.65\substack{+0.16\\-0.15}$ & $15.4\pm3.3$ & $1.74\substack{+0.47\\-0.42}$ & $329.6\substack{+3.8\\-3.7}$ & 0.785 & $+0.05\pm0.08$ & 1 & [17] \\
       HD 20329\,b & $1.72\pm0.07$ & $7.42\pm1.09$ & $8.06\pm1.53$ & $2141\pm27$ & 0.621 & -0.07$\pm$0.06 & 1 & [18] \\
       TOI-178\,b & $1.152\substack{+0.073\\-0.070}$ & $1.50\substack{+0.39\\-0.44}$ & $5.4\substack{+1.9\\-1.7}$ & $1040\substack{+22\\-21}$ & 0.545 & -0.23$\pm$0.05 & 6 & [19] \\
       TOI-178\,c & $1.669\substack{+0.114\\-0.099}$ & $4.77\substack{+0.55\\-0.68}$ & $5.62\substack{+1.50\\-1.30}$ & $873\pm18$ & 0.545 & -0.23$\pm$0.05 & 6 & [19] \\
       TOI-178\,f & $2.287\substack{+0.108\\-0.110}$ & $7.72\substack{+1.67\\-1.52}$ & $3.60\substack{+1.20\\-0.83}$ & $521\pm11$ & 0.545 & -0.23$\pm$0.05 & 6 & [19] \\
       HIP 9618\,b & $3.75\pm0.13$ & $8.40\pm2.00$ & $0.90\pm0.20$ & $685\pm17$ & 0.528 & -0.10$\pm$0.09 & 2 & [20,5]\\
       HIP 8152\,b & $2.54\substack{+0.16\\-0.14}$ & $8.90\pm1.70$ & $2.97\substack{+0.81\\-1.08}$ & $795\substack{+14\\-13}$ & 0.525 & -0.09$\pm$0.09 & 2 & [5,6] \\
       HIP 8152\,c & $2.52\substack{+0.15\\-0.14}$ & $10.7\substack{+2.1\\-2.2}$ & $3.63\substack{+1.02\\-1.38}$ & $651\pm11$ & 0.525 & -0.09$\pm$0.09 & 2 & [5,6] \\
    \hline
    \hline
    \end{tabular}
    \caption{Well-characterised planet below 4\,R$_\oplus$ orbiting kinematically thick disk stars. References: [1] \citet{Mortier_K2_111_2020}, [2] \citet{Bonomo_Kepler_K2_HARPSN_2023}, [3] \citet{Shporer_GJ1252_2020}, [4] \citet{Luque_GJ1252_2022}, [5] \citet{Akana_Murphy_TESS_Keck_2023}, [6] \citet{Polanski_TESS_Keck_2024}, [7] \citet{Peterson_Wolf503_2018}, [8] \citet{Mayo_K2_Campaign0_10_Planets_2018}, [9] \citet{Thygesen_K2_TESS_2023}, [10] \citet{Udry_HARPS_XLIV_2019}, [11] \citet{Delrez_HD136352_2021}, [12] \citet{Montalto_TOI5076_2024}, [13] \citet{Heidari_SOPHIE_TESS_2025}, [14] \citet{Batalha_Kepler10b_2011}, [15] \citet{Bonomo_Kepler10_2025}, [16] \citet{Fressin_Kepler10c_2011}, [17] \citet{Burt_TOI1231_2021}, [18] \citet{Murgas_HD20329_2022}, [19] \citet{Leleu_TOI178_2021}, [20] \citet{Osborn_HIP9618_2023}}
    \label{tab:thick_disk_planets}
    \end{center}
    
\end{table*}

\section{Conclusions}
\label{sec:Conclusion}
We present the discovery of a super-Earth and sub-Neptune orbiting the thick disk star TOI-2345. We detected these two planets using photometry from \textit{TESS} and \textit{CHEOPS}, and radial velocity observations from HARPS. We characterised the star to be slightly metal-poor and kinematically determined that TOI-2345 is a member of the Milky Way's thick disk. The photometric and spectroscopic observations allowed us to derive a radius of $1.504\substack{+0.047\\-0.044}$\,R$_\oplus$ and $2.451\substack{+0.045\\-0.046}$\,R$_\oplus$ and mass of $3.49\substack{+0.85\\-0.85}$\,M$_\oplus$ and $7.27\substack{+2.27\\-2.45}$\,M$_\oplus$ for the inner and outer planet orbiting the star at periods of 1.05\,d and $\sim$21\,d respectively. The accurate determination of the radii and masses of these planets enabled us to model the interior structure using plaNETic. To account for the connection in refractory elements between host stars and planets, we devolatise the host star with ExoInt and use the obtained abundances as priors in our interior modelling for the first time. This results in predicted interior structures which are in agreement to the interior structures obtained when using the stellar abundances as a prior. We obtained the interior structure of the inner rocky planet to have a large mantle, small core and nearly no atmosphere, while the outer planet also has a small core and large mantle but is gas-rich and has an atmospheric mass fraction of approximately 1\%. Additionally, performing atmospheric escape modelling we find that the Ultra-Short-Period inner super-Earth has lost its atmosphere, while the outer sub-Neptune has likely only lost a small fraction of its initial atmosphere. Finally, we place the TOI-2345 system in the context of thick disk stars. The characterisation of TOI-2345, which contains a USP super-Earth below the radius valley, demonstrates that such small planets around thick disk stars are more common than previously thought. Moreover, the orbital period of the planets infer that the USP underwent low-eccentricity migration. Since only a limited sample of small planets around thick disk stars has been well characterised to date, TOI-2345 adds onto this sample which will enable demographic studies of planets in this stellar population in the future.

\section*{Acknowledgements}
\textcolor{black}{We thank the anonymous reviewer for their feedback.}
\textit{CHEOPS} is an ESA mission in partnership with Switzerland with important contributions to the payload and the ground segment from Austria, Belgium, France, Germany, Hungary, Italy, Portugal, Spain, Sweden, and the United Kingdom. The \textit{CHEOPS} Consortium would like to gratefully acknowledge the support received by all the agencies, offices, universities, and industries involved. Their flexibility and willingness to explore new approaches were essential to the success of this mission. \textit{CHEOPS} data analysed in this article will be made available in the \textit{CHEOPS} mission archive (\url{https://cheops.unige.ch/archive_browser/}). 
Funding for the \textit{TESS} mission is provided by NASA's Science Mission Directorate. We acknowledge the use of public \textit{TESS} data from pipelines at the \textit{TESS} Science Office and at the \textit{TESS} Science Processing Operations Center. This research has made use of the Exoplanet Follow-up Observation Program website, which is operated by the California Institute of Technology, under contract with the National Aeronautics and Space Administration under the Exoplanet Exploration Program. 
YNEE acknowledges support from a Science and Technology Facilities Council (STFC) studentship, grant number ST/Y509693/1. 
TWi acknowledges support from the UKSA and the University of Warwick. 
S.G.S. acknowledge support from FCT through FCT contract nr. CEECIND/00826/2018 and POPH/FSE (EC). 
The Portuguese team thanks the Portuguese Space Agency for the provision of financial support in the framework of the PRODEX Programme of the European Space Agency (ESA) under contract number 4000142255. 
ML acknowledges support of the Swiss National Science Foundation under grant number PCEFP2\_194576. 
A. S. acknowledges support from the Swiss Space Office through the ESA PRODEX program. 
This work has been carried out within the framework of the NCCR PlanetS supported by the Swiss National Science Foundation under grants 51NF40\_182901 and 51NF40\_205606. 
A.J.M. acknowledges the support of the Swedish National Space Agency (grant 2023-00146).
HV acknowledges funding from the Institut Universitaire de France (IUF) that made this work possible. 
YAl acknowledges support from the Swiss National Science Foundation (SNSF) under grant 200020\_192038. 
DB, EP, EV, IR and RA acknowledge financial support from the Agencia Estatal de Investigación of the Ministerio de Ciencia e Innovación MCIN/AEI/10.13039/501100011033 and the ERDF “A way of making Europe” through projects PID2021-125627OB-C31, PID2021-125627OB-C32, PID2021-127289NB-I00, PID2023-150468NB-I00 and PID2023-149439NB-C41. 
from the Centre of Excellence “Severo Ochoa'' award to the Instituto de Astrofísica de Canarias (CEX2019-000920-S), the Centre of Excellence “María de Maeztu” award to the Institut de Ciències de l’Espai (CEX2020-001058-M), and from the Generalitat de Catalunya/CERCA programme. 
SCCB acknowledges the support from Fundação para a Ciência e Tecnologia (FCT) in the form of work contract through the Scientific Employment Incentive program with reference 2023.06687.CEECIND and DOI 10.54499/2023.06687.CEECIND/CP2839/CT0002. 
LBo, GBr, VNa, IPa, GPi, RRa, GSc, VSi, and TZi acknowledge support from \textit{CHEOPS} ASI-INAF agreement n. 2019-29-HH.0. 
ABr was supported by the SNSA. 
C.B. acknowledges support from the Swiss Space Office through the ESA PRODEX program. 
ACC acknowledges support from STFC consolidated grant number ST/V000861/1
and UKRI/ERC Synergy Grant EP/Z000181/1 (REVEAL).
ACMC acknowledges support from the FCT, Portugal, through the CFisUC projects UIDB/04564/2020 and UIDP/04564/2020, with DOI identifiers 10.54499/UIDB/04564/2020 and 10.54499/UIDP/04564/2020, respectively. 
A.C., A.D., B.E., K.G., and J.K. acknowledge their role as ESA-appointed \textit{CHEOPS} Science Team Members. 
P.E.C. is funded by the Austrian Science Fund (FWF) Erwin Schroedinger Fellowship, program J4595-N. 
This project was supported by the CNES. 
A.De. 
This work was supported by FCT - Funda\c{c}\~{a}o para a Ci\^{e}ncia e a Tecnologia through national funds and by FEDER through COMPETE2020 through the research grants UIDB/04434/2020, UIDP/04434/2020, 2022.06962.PTDC. 
O.D.S.D. is supported in the form of work contract (DL 57/2016/CP1364/CT0004) funded by national funds through FCT. 
B.-O. D. acknowledges support from the Swiss State Secretariat for Education, Research and Innovation (SERI) under contract number MB22.00046. 
A.C., A.D., B.E., K.G., and J.K. acknowledge their role as ESA-appointed \textit{CHEOPS} Science Team Members. 
This project has received funding from the Swiss National Science Foundation for project 200021\_200726. It has also been carried out within the framework of the National Centre of Competence in Research PlanetS supported by the Swiss National Science Foundation under grant 51NF40\_205606. The authors acknowledge the financial support of the SNSF. 
MF and CMP gratefully acknowledge the support of the Swedish National Space Agency (DNR 65/19, 174/18). 
DG gratefully acknowledges financial support from the CRT foundation under Grant No. 2018.2323 “Gaseousor rocky? Unveiling the nature of small worlds”. 
M.G. is an F.R.S.-FNRS Senior Research Associate. 
MNG is the ESA \textit{CHEOPS} Project Scientist and Mission Representative. BMM is the ESA \textit{CHEOPS} Project Scientist. KGI was the ESA \textit{CHEOPS} Project Scientist until the end of December 2022 and Mission Representative until the end of January 2023. All of them are/were responsible for the Guest Observers (GO) Programme. None of them relay/relayed proprietary information between the GO and Guaranteed Time Observation (GTO) Programmes, nor do/did they decide on the definition and target selection of the GTO Programme. 
CHe acknowledges financial support from the Österreichische Akademie 1158 der Wissenschaften and from the European Union H2020-MSCA-ITN-2019 1159 under Grant Agreement no. 860470 (CHAMELEON). Calculations were performed using supercomputer resources provided by the Vienna Scientific Cluster (VSC). 
A.C., A.D., B.E., K.G., and J.K. acknowledge their role as ESA-appointed \textit{CHEOPS} Science Team Members. The authors acknowledge support from the Swiss NCCR PlanetS and the Swiss National Science Foundation. This work has been carried out within the framework of the NCCR PlanetS supported by the Swiss National Science Foundation under grants 51NF40182901 and 51NF40205606. J.K. acknowledges support from the Swiss National Science Foundation under grant number TMSGI2$\_$211697. 
K.W.F.L. was supported by Deutsche Forschungsgemeinschaft grants RA714/14-1 within the DFG Schwerpunkt SPP 1992, Exploring the Diversity of Extrasolar Planets. 
This work was granted access to the HPC resources of MesoPSL financed by the Region Ile de France and the project Equip@Meso (reference ANR-10-EQPX-29-01) of the programme Investissements d'Avenir supervised by the Agence Nationale pour la Recherche. 
This work has been carried out within the framework of the NCCR PlanetS supported by the Swiss National Science Foundation under grants 51NF40\_182901 and 51NF40\_205606. AL acknowledges support of the Swiss National Science Foundation under grant number  TMSGI2\_211697. 
PM acknowledges support from STFC research grant number ST/R000638/1. 
This work was also partially supported by a grant from the Simons Foundation (PI Queloz, grant number 327127). 
NCS is co-funded by the European Union (ERC, FIERCE, 101052347). Views and opinions expressed are however those of the author(s) only and do not necessarily reflect those of the European Union or the European Research Council. Neither the European Union nor the granting authority can be held responsible for them. This work was supported by FCT - Fundação para a Ciência e a Tecnologia through national funds by these grants: UIDB/04434/2020 DOI: 10.54499/UIDB/04434/2020, UIDP/04434/2020 DOI: 10.54499/UIDP/04434/2020.
GyMSz acknowledges the support of the Hungarian National Research, Development and Innovation Office (NKFIH) grant K-125015, a a PRODEX Experiment Agreement No. 4000137122, the Lendület LP2018-7/2021 grant of the Hungarian Academy of Science and the support of the city of Szombathely. 
V.V.G. is an F.R.S-FNRS Research Associate. 
JV acknowledges support from the Swiss National Science Foundation (SNSF) under grant PZ00P2$\_$208945. 
NAW acknowledges UKSA grant ST/R004838/1.

\section*{Data Availability}
The data underlying this article will be made available in the \textit{CHEOPS} mission archive \href{https://cheops.unige.ch/archive_browser/}{https://cheops.unige.ch/archive$\_$browser/}. This paper includes data collected by the \textit{TESS} mission, which is publicly available from the Mikulski Archive for Space Telescopes (MAST) at the Space Telescope Science Institute (STScI) \href{https://mast.stsci.edu}{https://mast.stsci.edu}.



\bibliographystyle{mnras}
\bibliography{example} 

\vspace{1cm}
\noindent 
$^{1}$Department of Physics, University of Warwick, Gibbet Hill Road, Coventry CV4 7AL, United Kingdom \\
$^{2}$Space Research Institute, Austrian Academy of Sciences, Schmiedlstrasse 6, A-8042 Graz, Austria \\
$^{3}$Department of Space, Earth and Environment, Chalmers University of Technology, Onsala Space Observatory, 439 92 Onsala, Sweden \\
$^{4}$Instituto de Astrofisica e Ciencias do Espaco, Universidade do Porto, CAUP, Rua das Estrelas, 4150-762 Porto, Portugal \\
$^{5}$Observatoire astronomique de l'Université de Genève, Chemin Pegasi 51, 1290 Versoix, Switzerland \\
$^{6}$Space Research and Planetary Sciences, Physics Institute, University of Bern, Gesellschaftsstrasse 6, 3012 Bern, Switzerland \\
$^{7}$Center for Space and Habitability, University of Bern, Gesellschaftsstrasse 6, 3012 Bern, Switzerland \\
$^{8}$Department of Astronomy, Stockholm University, AlbaNova University Center, 10691 Stockholm, Sweden \\
$^{9}$Lund Observatory, Division of Astrophysics, Department of Physics, Lund University, Box 118, 221 00, Lund, Sweden \\
$^{10}$ETH Zurich, Department of Physics, Wolfgang-Pauli-Strasse 2, CH-8093 Zurich, Switzerland \\
$^{11}$Aix Marseille Univ, CNRS, CNES, LAM, 38 rue Frédéric Joliot-Curie, 13388 Marseille, France \\
$^{12}$Instituto de Astrofísica de Canarias, Vía Láctea s/n, 38200 La Laguna, Tenerife, Spain \\
$^{13}$Departamento de Astrofísica, Universidad de La Laguna, Astrofísico Francisco Sanchez s/n, 38206 La Laguna, Tenerife, Spain \\
$^{14}$Admatis, 5. Kandó Kálmán Street, 3534 Miskolc, Hungary \\
$^{15}$Depto. de Astrofísica, Centro de Astrobiología (CSIC-INTA), ESAC campus, 28692 Villanueva de la Cañada (Madrid), Spain \\
$^{16}$Departamento de Fisica e Astronomia, Faculdade de Ciencias, Universidade do Porto, Rua do Campo Alegre, 4169-007 Porto, Portugal \\
$^{17}$INAF, Osservatorio Astronomico di Padova, Vicolo dell'Osservatorio 5, 35122 Padova, Italy \\
$^{18}$Institute of Space Research, German Aerospace Center (DLR), Rutherfordstrasse 2, 12489 Berlin, Germany \\
$^{19}$SETI Institute, Mountain View, CA 94043 USA/NASA Ames Research Center, Moffett Field, CA 94035 USA \\
$^{20}$Centre for Exoplanet Science, SUPA School of Physics and Astronomy, University of St Andrews, North Haugh, St Andrews KY16 9SS, UK \\
$^{21}$CFisUC, Departamento de Física, Universidade de Coimbra, 3004-516 Coimbra, Portugal \\
$^{22}$INAF, Osservatorio Astrofisico di Torino, Via Osservatorio, 20, I-10025 Pino Torinese To, Italy \\
$^{23}$Centre for Mathematical Sciences, Lund University, Box 118, 221 00 Lund, Sweden \\
$^{24}$ARTORG Center for Biomedical Engineering Research, University of Bern, Bern, Switzerland \\
$^{25}$ELTE Gothard Astrophysical Observatory, 9700 Szombathely, Szent Imre h. u. 112, Hungary \\
$^{26}$SRON Netherlands Institute for Space Research, Niels Bohrweg 4, 2333 CA Leiden, Netherlands \\
$^{27}$Centre Vie dans l’Univers, Faculté des sciences, Université de Genève, Quai Ernest-Ansermet 30, 1211 Genève 4, Switzerland \\
$^{28}$Leiden Observatory, University of Leiden, PO Box 9513, 2300 RA Leiden, The Netherlands \\
$^{29}$Dipartimento di Fisica, Università degli Studi di Torino, via Pietro Giuria 1, I-10125, Torino, Italy \\
$^{30}$National and Kapodistrian University of Athens, Department of Physics, University Campus, Zografos GR-157 84, Athens, Greece \\
$^{31}$Astrobiology Research Unit, Université de Liège, Allée du 6 Août 19C, B-4000 Liège, Belgium \\
$^{32}$Kavli Institute for Astrophysics and Space Research, Massachusetts Institute of Technology, Cambridge, MA, USA \\
$^{33}$Department of Astrophysics, University of Vienna, Türkenschanzstrasse 17, 1180 Vienna, Austria \\
$^{34}$European Space Agency (ESA), European Space Research and Technology Centre (ESTEC), Keplerlaan 1, 2201 AZ Noordwijk, The Netherlands \\
$^{35}$Institute for Theoretical Physics and Computational Physics, Graz University of Technology, Petersgasse 16, 8010 Graz, Austria \\
$^{36}$NASA Ames Research Center, Moﬀett Field, CA 94035, USA \\
$^{37}$Konkoly Observatory, Research Centre for Astronomy and Earth Sciences, 1121 Budapest, Konkoly Thege Miklós út 15-17, Hungary \\
$^{38}$ELTE E\"otv\"os Lor\'and University, Institute of Physics, P\'azm\'any P\'eter s\'et\'any 1/A, 1117 Budapest, Hungary \\
$^{39}$IMCCE, UMR8028 CNRS, Observatoire de Paris, PSL Univ., Sorbonne Univ., 77 av. Denfert-Rochereau, 75014 Paris, France \\
$^{40}$Institut d'astrophysique de Paris, UMR7095 CNRS, Université Pierre \& Marie Curie, 98bis blvd. Arago, 75014 Paris, France \\
$^{41}$Astrophysics Group, Lennard Jones Building, Keele University, Staffordshire, ST5 5BG, United Kingdom \\
$^{42}$European Space Agency, ESA - European Space Astronomy Centre, Camino Bajo del Castillo s/n, 28692 Villanueva de la Cañada, Madrid, Spain \\
$^{43}$INAF, Osservatorio Astrofisico di Catania, Via S. Sofia 78, 95123 Catania, Italy \\
$^{44}$Weltraumforschung und Planetologie, Physikalisches Institut, University of Bern, Gesellschaftsstrasse 6, 3012 Bern, Switzerland \\
$^{45}$Dipartimento di Fisica e Astronomia "Galileo Galilei", Università degli Studi di Padova, Vicolo dell'Osservatorio 3, 35122 Padova, Italy \\
$^{46}$Cavendish Laboratory, JJ Thomson Avenue, Cambridge CB3 0HE, UK \\
$^{47}$German Aerospace Center (DLR), Markgrafenstrasse 37, 10117 Berlin, Germany \\
$^{48}$Institut fuer Geologische Wissenschaften, Freie Universitaet Berlin, Malteserstrasse 74-100,12249 Berlin, Germany \\
$^{49}$Institut de Ciencies de l'Espai (ICE, CSIC), Campus UAB, Can Magrans s/n, 08193 Bellaterra, Spain \\
$^{50}$Institut d'Estudis Espacials de Catalunya (IEEC), 08860 Castelldefels (Barcelona), Spain \\
$^{51}$Space sciences, Technologies and Astrophysics Research (STAR) Institute, Université de Liège, Allée du 6 Août 19C, 4000 Liège, Belgium \\
$^{52}$HUN-REN-ELTE Exoplanet Research Group, Szent Imre h. u. 112., Szombathely, H-9700, Hungary \\
$^{53}$Leiden Observatory, University of Leiden, Einsteinweg 55, 2333 CA Leiden, The Netherlands \\
$^{54}$Institute of Astronomy, University of Cambridge, Madingley Road, Cambridge, CB3 0HA, United Kingdom \\
$^{55}$Planetary Discoveries, Valencia, CA 91354, USA \\
$^{56}$Department of Physics, Engineering and Astronomy, Stephen F. Austin State University, 1936 North St, Nacogdoches, TX 75962, USA

\appendix

\textcolor{black}{\section{\textit{TESS} Photometry}}

\begin{figure*}
    \centering
    \includegraphics[width=\linewidth]{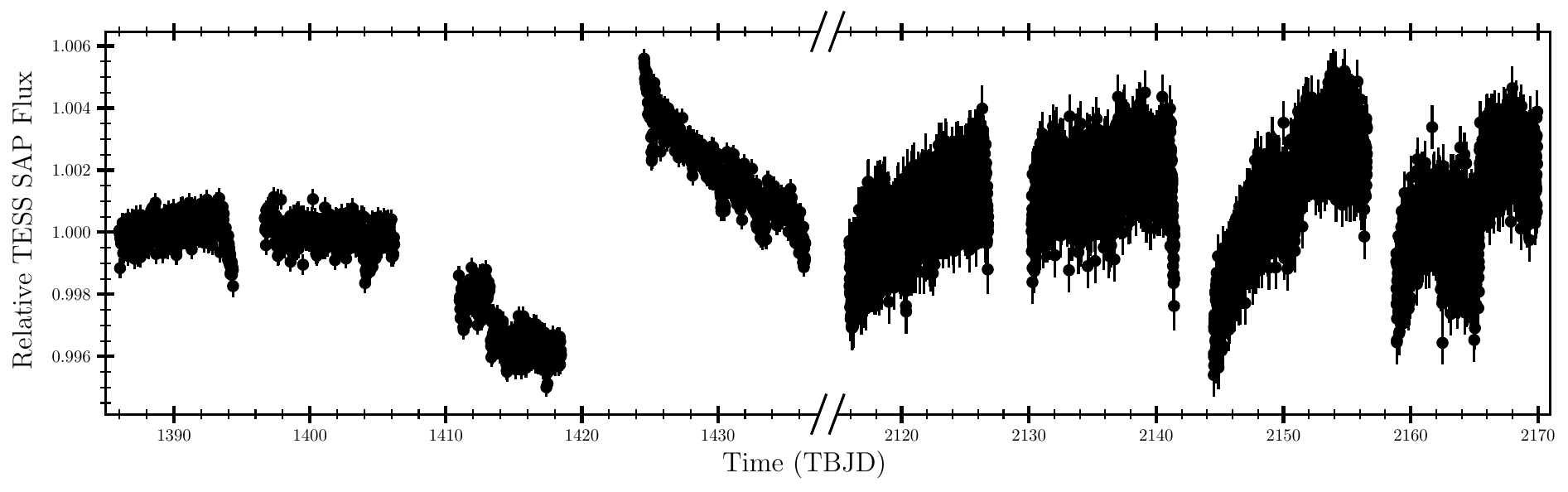}
    \caption{\textcolor{black}{TESS SAP flux of TOI-2345 in sectors 3, 4, 30 and 31.}}
    \label{fig:tess_sap}
\end{figure*}

\begin{figure*}
    \centering
    \includegraphics[width=\linewidth]{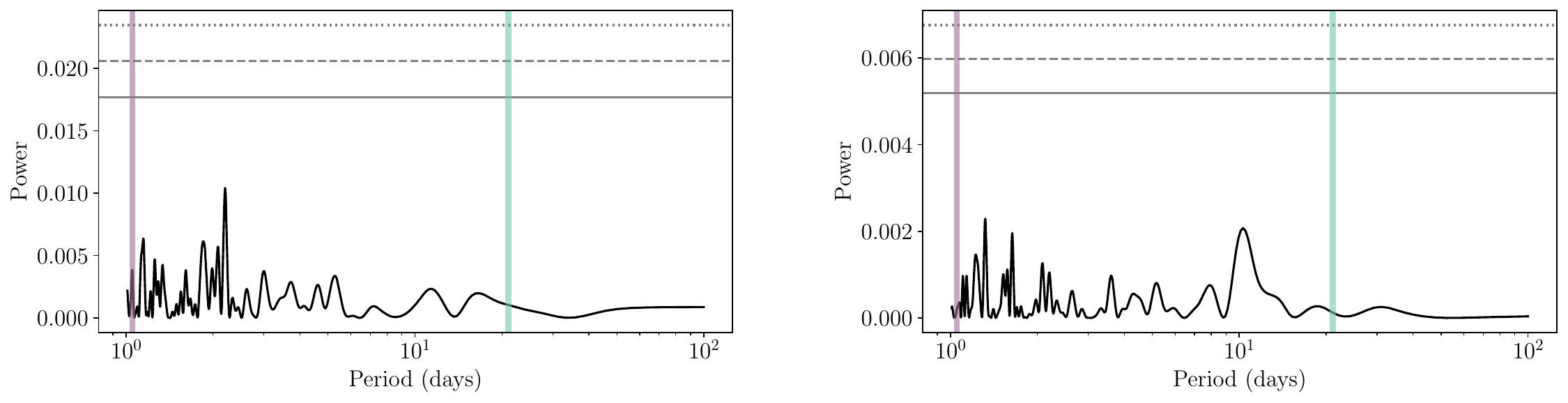}
    \caption{\textcolor{black}{Lomb-Scargle-periodogram of the available TESS data of sectors 3 \& 4 (left) and sectors 30 \& 31 (right). The orbital periods of TOI-2345\,b and TOI-2345\,c are shown by the purple and green line respectively.}}
    \label{fig:tess_periodogram}
\end{figure*}

\begin{figure*}
    \centering
    \includegraphics[width=\linewidth]{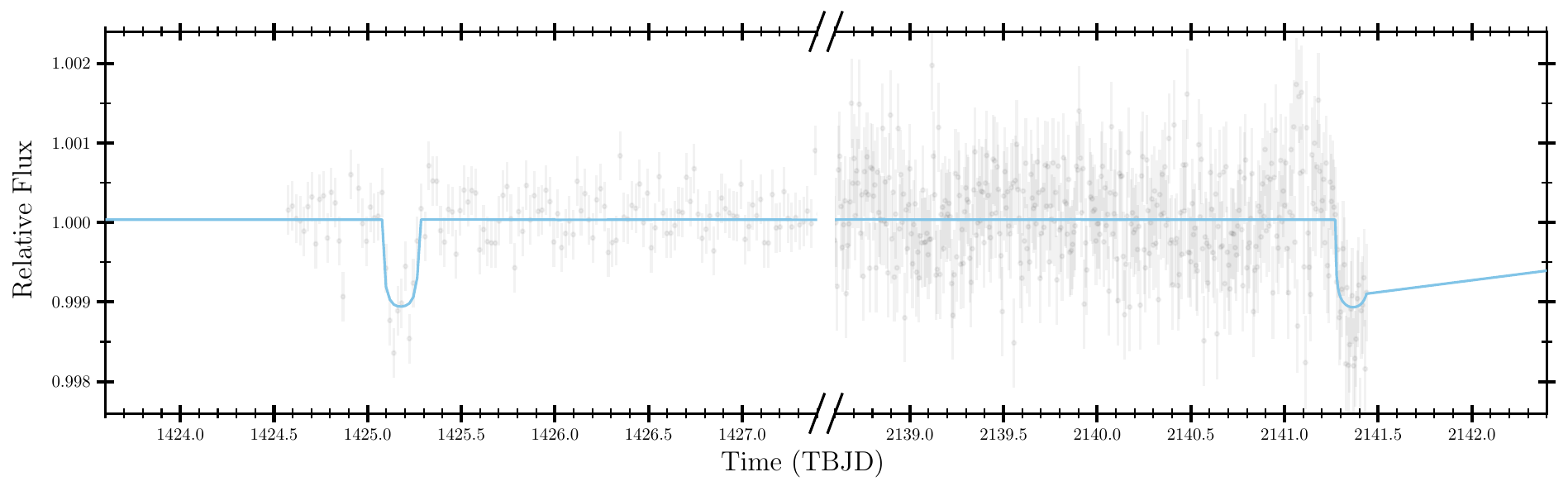}
    \caption{\textcolor{black}{TESS photometry and our best fit model in juliet of the even transits of TOI-2345\,c (transit 2 and 4) that are close to the sector gaps and were missed in the TESS vetting process. The  signals of planet b are removed and the model is only showing planet c.}}
    \label{fig:tess_even_transits}
\end{figure*}

\section{\textit{CHEOPS} Photometry}

\begin{table*}
    \caption{Selected \textit{CHEOPS} detrending vectors for DRP and PIPE of each visit. The detrending vectors are abbreviated as time (t), roll angle ($\phi$), background (bg), and centroid offset (x, y).}
    \centering
    \begin{tabular}{cccc}
    \hline
    \hline
        Visit & Aperture & DRP & PIPE \\
        &  (pixel) & Detrending & Detrending\\
        \hline
        1 & R16 & bg, t, cos$\phi$ & y, cos2$\phi$, sin3$\phi$, cos3$\phi$\\
        2 & R17 & bg, sin$\phi$, y$^2$, cos$\phi$, sin2$\phi$, y, cos2$\phi$ & cos2$\phi$, y, x$^2$, x, cos3$\phi$\\
        3 & R17 & bg, x, x$^2$ & y, cos2$\phi$, bg, x, cos$\phi$, sin3$\phi$\\
        4 & R16 & bg, t, sin$\phi$, x & sin$\phi$, cos2$\phi$, bg cos3$\phi$\\
        5 & R16 & bg, x, sin$\phi$ & sin$\phi$, y, sin3$\phi$, cos2$\phi$, x$^2$, t\\
    \hline
    \hline
    \end{tabular}
    \label{tab:CHEOPS_detrending}
\end{table*}

\begin{figure*}
    \centering
    \includegraphics[width=0.9\columnwidth]{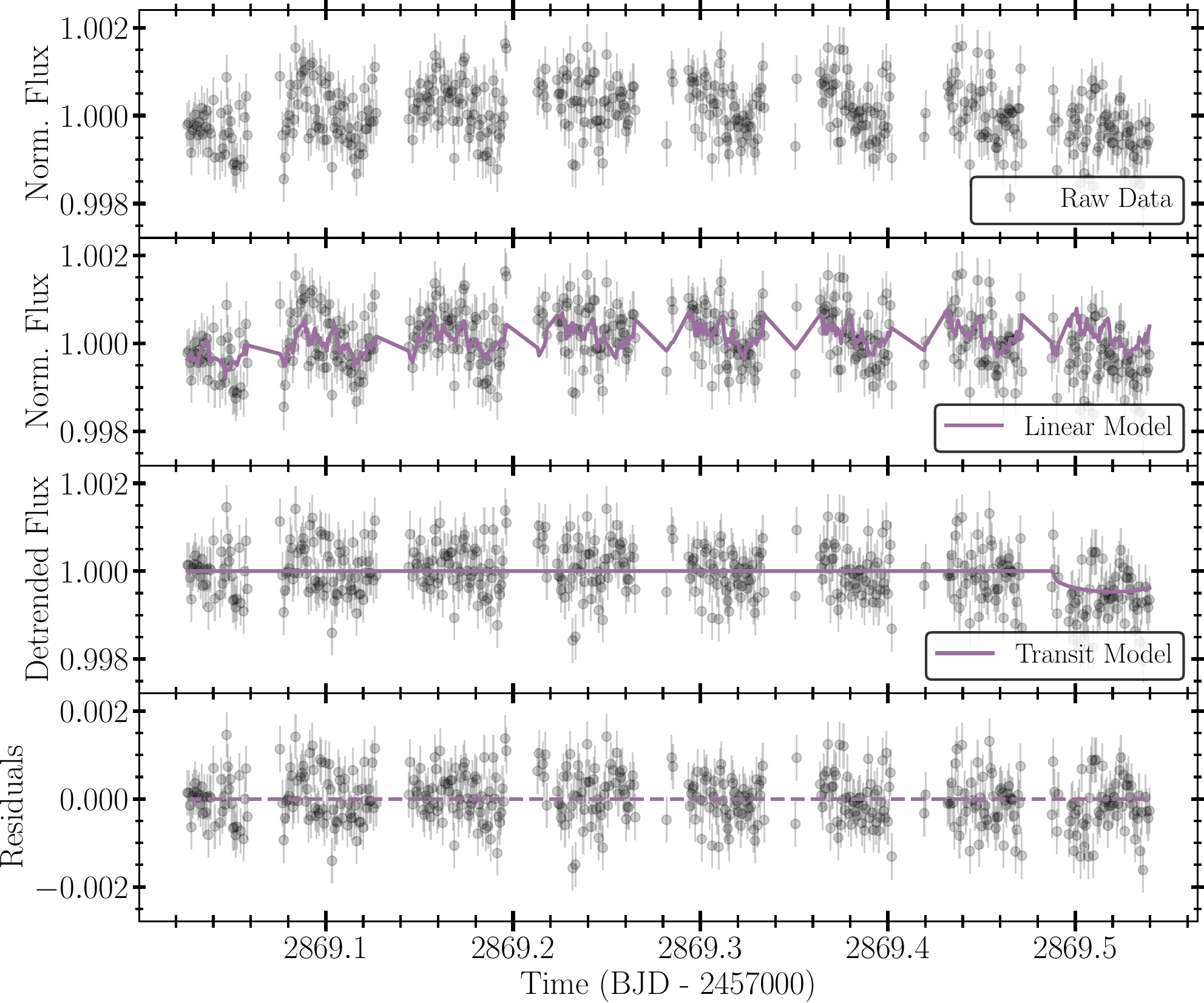}
    \qquad
    \includegraphics[width=0.9\columnwidth]{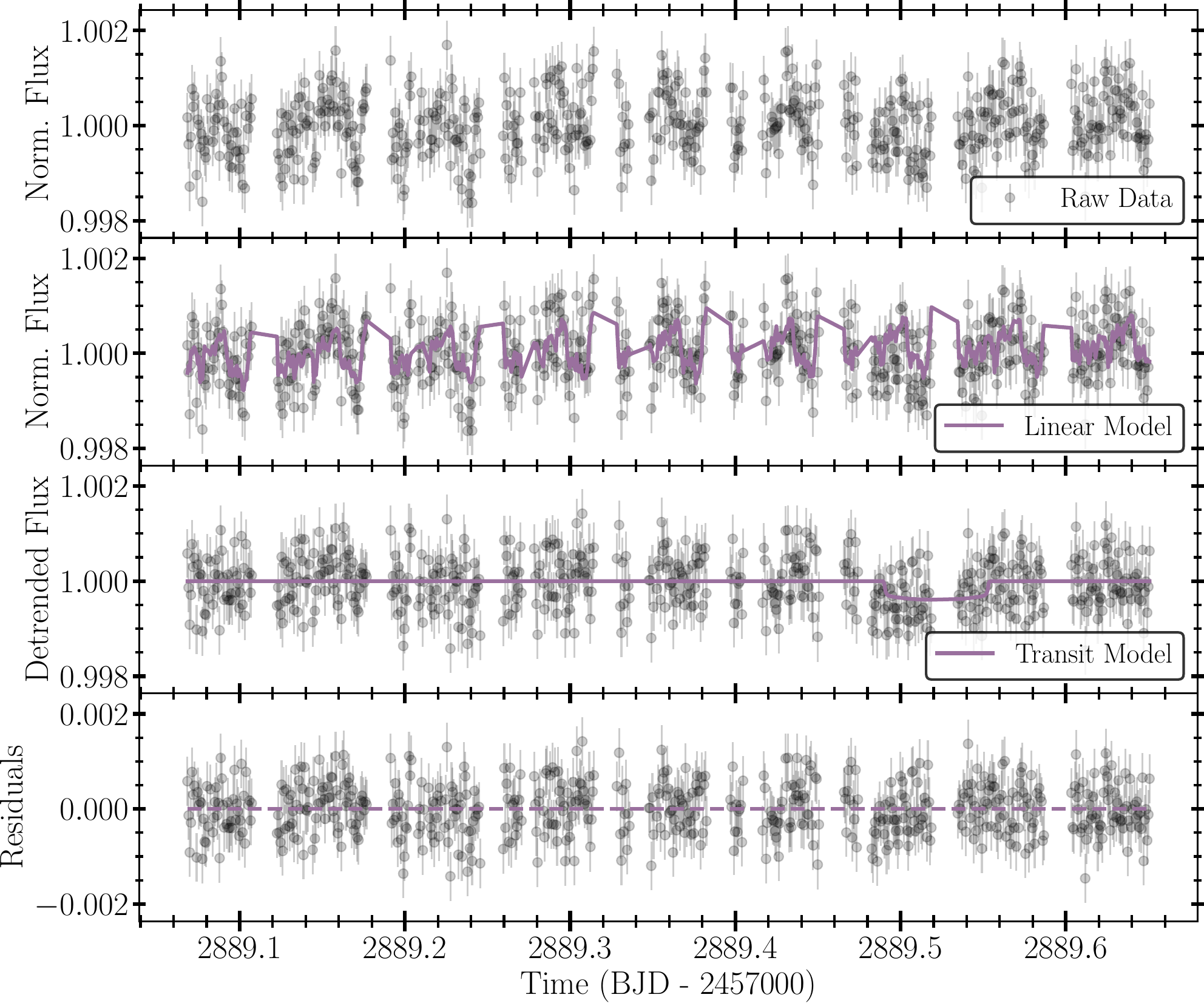}
    \qquad
    \includegraphics[width=0.9\columnwidth]{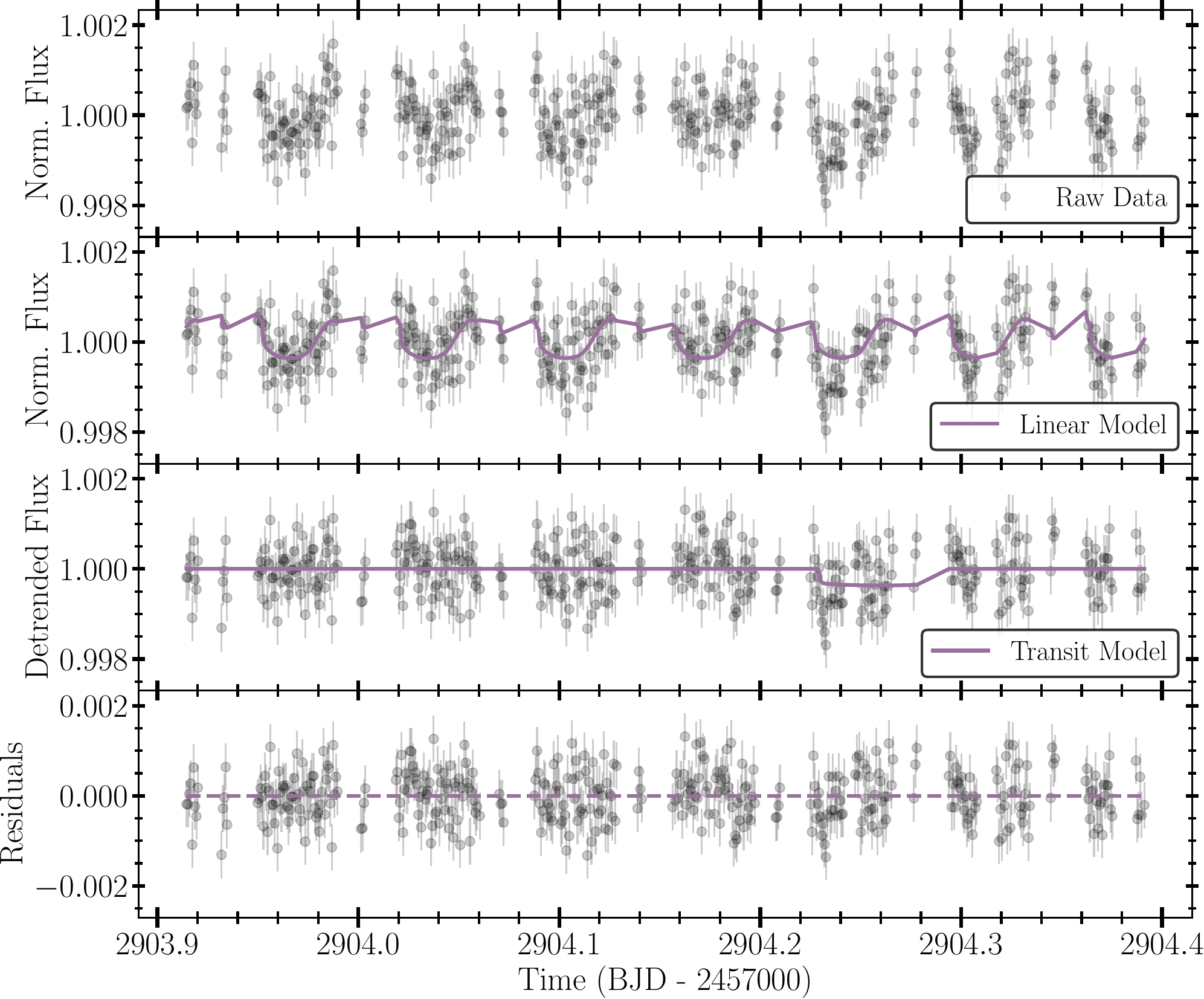}
    \qquad
    \includegraphics[width=0.9\columnwidth]{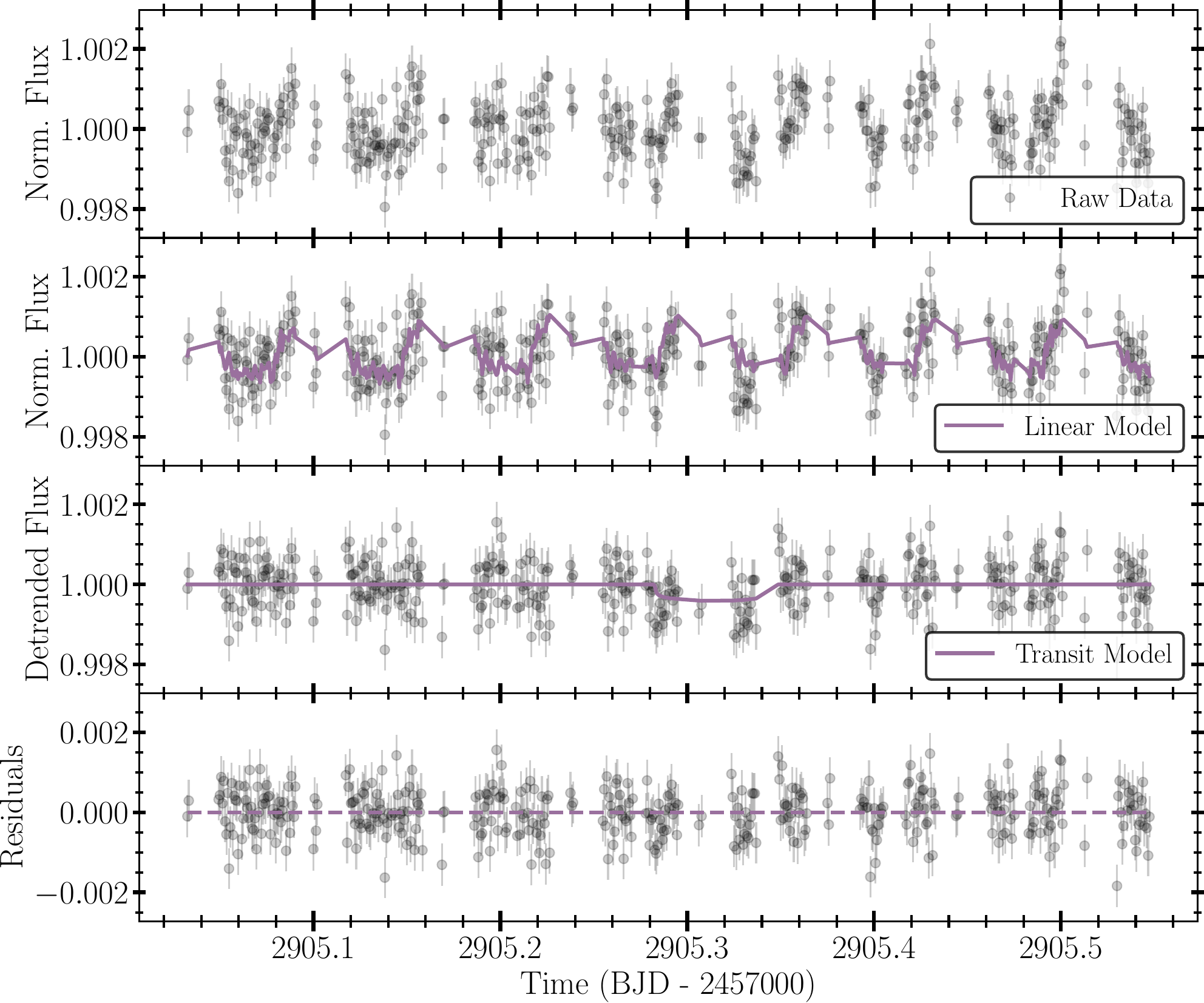}
    \qquad
    \includegraphics[width=0.9\columnwidth]{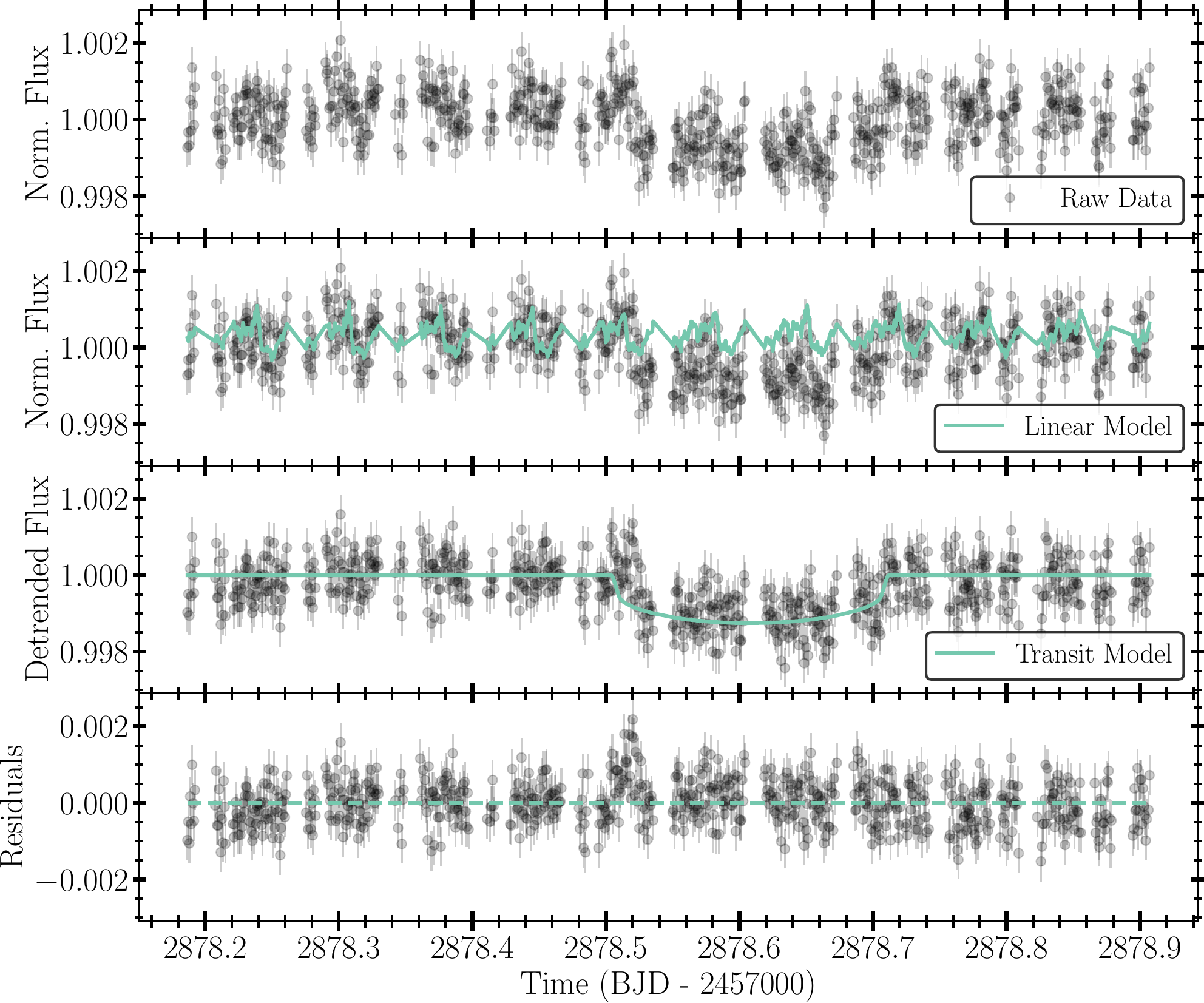}
    \caption{\textit{CHEOPS} visits 1001, 1002, 3201, 3202 of planet b and 2401 of planet c and their detrending. The top row shows the data, the second row the linear model based on the selected detrending vectors, the third row shows the transit model and at the bottom the residuals of the fit are presented. }
    \label{fig:cheops_detrending_plots}
\end{figure*}

\textcolor{black}{\section{HARPS Radial Velocities}}

\begin{figure*}
    \centering
    \includegraphics[width=\linewidth]{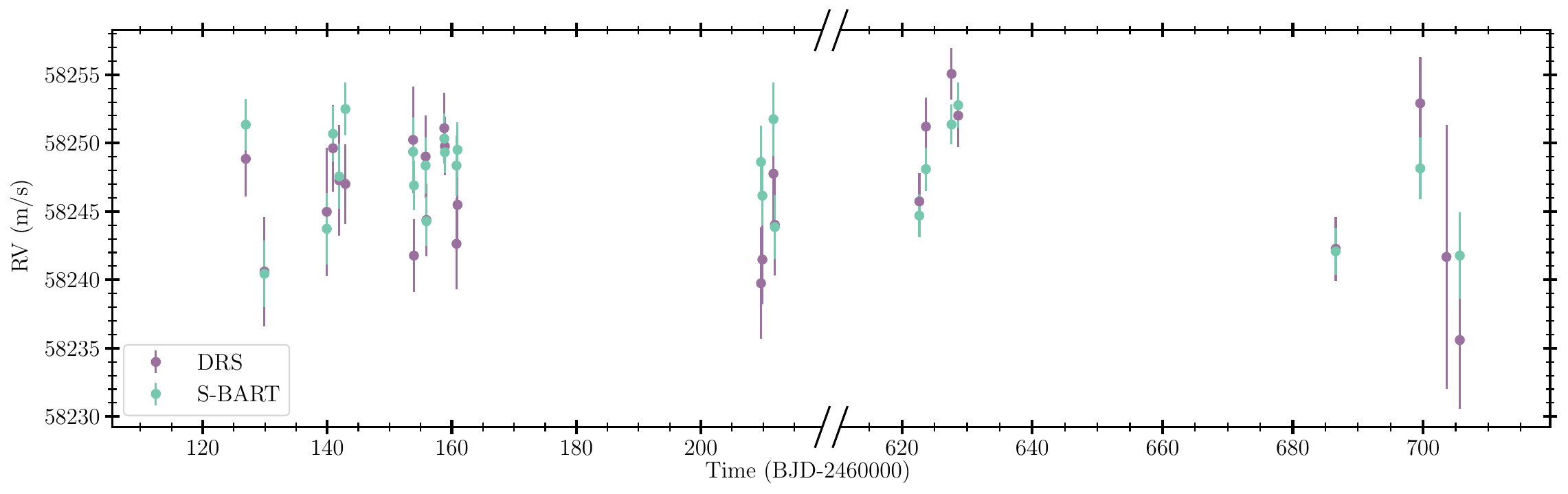}
    \caption{\textcolor{black}{HARPS RVs as extracted from the DRS (purple) and \textsc{s-bart} (green) taken between 2023/07/01 and 2025/01/30.}}
    \label{fig:rv_timeseries}
\end{figure*}



\begin{table*}
\caption{\textcolor{black}{Radial Velocities from the HARPS DRS and the \textsc{s-bart} extraction.}}
    \centering
    \begin{tabular}{|c|c|c|c|c|}
    \hline \hline
        Time (RJD) & DRS RV (m/s) & $\sigma_\text{DRS RV}$ (m/s) & \textsc{s-bart} RV (m/s) & $\sigma_\text{\textsc{s-bart} RV}$ (m/s) \\ \hline

        60126.93 & 58248.8 & 2.8 & 58251.3 & 1.9  \\
        60129.92 & 58240.6 & 4.0 & 58240.4 & 2.4\\ 
        60139.93 & 58245.0 & 4.7 & 58243.7 & 2.6  \\ 
        ... & ... & ... & ... & ... \\\hline \hline
    \end{tabular}
\end{table*}

\begin{table*}
\caption{\textcolor{black}{Activity Indicators from the HARPS DRS.}}
    \centering
    \begin{tabular}{|c|c|c|c|c|c|c|c|c|c|c|c|c|c|}
    \hline \hline
        Time (RJD) & FWHM (m/s) & BIS (m/s) &  Contrast & S-Index & H$\alpha$  & Na & Ca  \\ \hline

        60126.93 & 6620.9$\pm$5.5 & -32.3$\pm$5.5 & 65.877$\pm$0.055 & 195.5$\pm$4.3 & 0.25766$\pm$0.00045 & 0.13288$\pm$0.00031 & 0.1576$\pm$0.0041  \\
        60129.92 & 6618.2$\pm$8.0 & -14.8$\pm$8.0 & 66.166$\pm$0.080 & 174.0$\pm$7.1 & 0.25955$\pm$0.00069 & 0.13122$\pm$0.00046 & 0.1369$\pm$0.0068  \\ 
        60139.93 & 6619.0$\pm$9.4 & -25.7$\pm$9.5 & 65.888$\pm$0.093 & 182.0$\pm$7.9 & 0.25758$\pm$0.00083 & 0.13309$\pm$0.00056 & 0.1446$\pm$0.0076  \\ 
        ... & ... & ... & ... & ... & ... & ... & ...  \\ \hline \hline
    \end{tabular}
\end{table*}

\section{Long-term Photometry}

\begin{figure*}
    \centering
    \includegraphics[width=\linewidth]{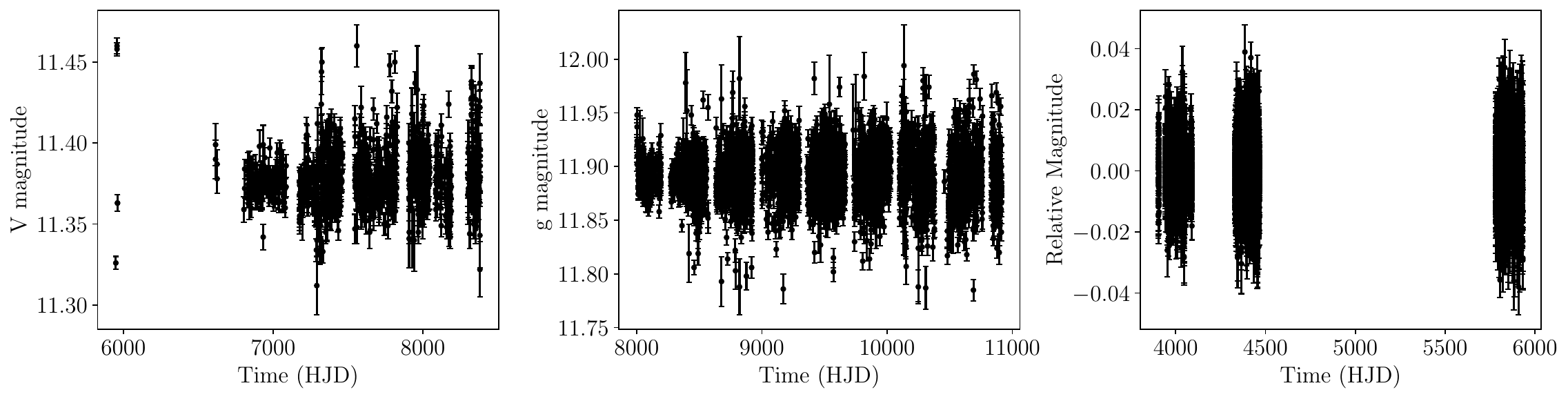}
    \caption{Lightcurves from ASAS-SN \textcolor{black}{V-Band (left) and ASAS-SN g-Band (middle)} and WASP (right) of TOI-2345.}
    \label{fig:asas_sn_wasp_lightcurves}
\end{figure*}

\section{Priors and Fitted Parameters}
\begin{table*}
    \caption{Stellar and Instrumental Parameters of the joint fit in \textcolor{black}{\textsc{juliet}}. Uniform distribustions are noted by $\mathcal{U}$, normal distributions by $\mathcal{N}$ and Log-uniform distributions by $\mathcal{L}$.}
    \centering
    \begin{tabular}{lccc}
    \hline \hline
        Parameter & Unit & Prior & Posterior  \\
        \hline
        $\rho_*$ & (kg/m$^3$) & $\mathcal{N}\textcolor{black}{(2645,142)}$ & $\textcolor{black}{2172\substack{+57\\-63}}$ \\
        GP$_\rho$ \textit{TESS} & (d) & $\mathcal{L}(0.001,1000)$ & $\textcolor{black}{21\substack{+291\\-21}}$ \\
        GP$_\sigma$ \textit{TESS} & (ppm) & $\mathcal{L}(0.000001,1000000)$ & $0.0000098\substack{+0.0000363\\-0.0000069}$\\
        q$_1$ \textit{TESS} & - & $\mathcal{U}(0.0,1.0)$ & $0.31\substack{+0.19\\-0.17}$\\
        q$_2$ \textit{TESS} & - & $\mathcal{U}(0.0,1.0)$ & $0.20\substack{+0.29\\-0.14}$\\
        q$_1$ \textit{CHEOPS} & - & $\mathcal{U}(0.0,1.0)$ & $0.106\substack{+0.092\\-0.058}$\\
        q$_2$ \textit{CHEOPS} & - & $\mathcal{U}(0.0,1.0)$ & $0.63\substack{+0.24\\-0.30}$\\
        jitter \textit{TESS} & (ppm) & $\mathcal{L}(0.1,1000)$ & $3.57\substack{+31.28\\-3.25}$\\
        jitter \textit{CHEOPS} & (ppm) & $\mathcal{L}(0.1,100000)$ & $193.60\substack{+21.58\\-22.62}$\\
        offset \textit{TESS} & - & $\mathcal{N}(0,0.1)$ & $-0.0000337\substack{+0.0000061\\-0.0000065}$\\
        offset \textit{CHEOPS} & - & $\mathcal{N}(0,0.1)$ & $0.000017\substack{+0.000010\\-0.000010}$\\
        jitter HARPS & (m/s) & $\mathcal{L}(0.001,100)$ & $1.69\substack{+0.56\\-0.52}$\\
        offset HARPS & (m/s) & $\mathcal{U}(58200,59300)$ & $58248.06\substack{+0.51\\-0.49}$\\
        \hline \hline
    \end{tabular}
    \label{tab:instr_fit}
\end{table*}

\bsp	
\label{lastpage}
\end{document}